\documentclass[a4paper,12pt]{article}
\usepackage{amssymb}
\usepackage[dvips]{graphicx}
\usepackage[usenames]{color}
\usepackage{amsmath}
\usepackage{fullpage}
\usepackage{boxedminipage}
\usepackage{listings}
\usepackage{minitoc}
\usepackage{wrapfig}

\input epsf
\newcommand{\be}{\begin{equation}}
\newcommand{\ee}{\end{equation}}
\newcommand{\bea}{\begin{eqnarray}}
\newcommand{\eea}{\end{eqnarray}}
\newcommand{\ba}{\begin{array}}
\newcommand{\ea}{\end{array}}

\renewcommand{\Im}{{\tt Im\,}}
\renewcommand{\Re}{{\tt Re\,}}

\begin{document}

\begin{titlepage}

    \thispagestyle{empty}
    \begin{flushright}
        \hfill{CERN-PH-TH/180} \\\hfill{UCLA/07/TEP/20}\\
    \end{flushright}

    \vspace{5pt}
    \begin{center}
        { \LARGE{\textbf{Separation of Attractors\\\vskip.5cm
        in 1-modulus Quantum Corrected\\\vskip.5cm Special Geometry}}}\vspace{15pt}
        \vspace{35pt}

        { {\textbf{S.Bellucci$^{1}$, S.Ferrara$^{2,1,3}$, A.Marrani$^{4,1}$ and A.Shcherbakov$^{1}$}}}\vspace{15pt}

         {$1$ \it INFN - Laboratori Nazionali di Frascati, \\
        Via Enrico Fermi 40,00044 Frascati, Italy\\
        \texttt{bellucci, marrani, ashcherb@lnf.infn.it}}

        \vspace{10pt}

        {$2$ \it Physics Department,Theory Unit, CERN, \\
        CH 1211, Geneva 23, Switzerland\\
        \texttt{sergio.ferrara@cern.ch}}

        \vspace{10pt}

         {$3$ \it Department of Physics and Astronomy,\\
        University of California, Los Angeles, CA USA\\
        \texttt{ferrara@physics.ucla.edu}}

         \vspace{10pt}

        {$4$ \it Museo Storico della Fisica e\\
        Centro Studi e Ricerche ``Enrico Fermi"\\
        Via Panisperna 89A, 00184 Roma, Italy}

        \vspace{15pt}
\end{center}

\begin{abstract}
We study the attractor equations for a quantum corrected
prepotential $\mathcal{F}=t^{3}+i\lambda $, with $%
\lambda \in \mathbb{R}$,which is the only correction which preserves
the axion shift symmetry and modifies the geometry.

By performing computations in the ``magnetic'' charge configuration,
we find evidence for interesting phenomena (absent in the classical
limit of vanishing $\lambda $). For a certain range of the quantum
parameter $\lambda $ we find a \textit{``separation''} of
attractors, \textit{i.e.} the existence of multiple solutions to the
Attractor Equations for fixed supporting charge configuration.
Furthermore, we find that, away from the classical limit, a
\textit{``transmutation''} of the supersymmetry-preserving features
of the attractors takes place when $\lambda $ reaches a particular
critical value.
\end{abstract}

\end{titlepage}

\section{\label{Intro}Introduction}

After the discovery of the \textit{Attractor Mechanism} in the mid 90's \cite
{FKS}-\nocite{Strom,FK1,FK2}\cite{FGK} in the context of BPS black holes
(BHs), recently \textit{extremal BH attractors} have been object of
intensive study \cite{Sen-old1}--\nocite
{GIJT,Sen-old2,K1,TT,G,GJMT,Ebra1,K2,Ira1,Tom,
BFM,AoB-book,FKlast,Ebra2,BFGM1,rotating-attr,K3,Misra1,Lust2,BFMY,CdWMa,
DFT07-1,BFM-SIGRAV06,Cer-Dal,ADFT-2,Saraikin-Vafa-1,Ferrara-Marrani-1,TT2,
ADOT-1,fm07,CCDOP,Misra2,Astefanesei,Anber,Myung1,CFM1,BMOS-1,Hotta,Gao,
PASCOS07,Sen-review,Belhaj1}\cite{AFMT1}. This is mainly due to the
(re)discovery of new classes of scalar attractor configurations, which do
not saturate the BPS bound and, when considering a supergravity theory,
break all supersymmetries at the BH event horizon.

The various classes of BPS and non-BPS attractors are strictly related to
the geometry of the scalar manifold. In $\mathcal{N}=2$, $d=4$ ungauged
supergravity coupled to $n_{V}$ Abelian vector multiplets, the scalar
manifold is endowed with the so-called Special K\"{a}hler (SK) geometry (see
\textit{e.g.} \cite{CDF-review} and Refs. therein), in which the holomorphic
prepotential function $\mathcal{F}$ plays a key role. In general, SK
geometry admits three classes of extremal BH attractors: $\frac{1}{2}$-BPS
(preserving four supersymmetries out of the eight pertaining to asymptotical
$\mathcal{N}=2$, $d=4$ superPoincar\'{e} algebra, and two non-supersymmetric
typologies, discriminated by the eventual vanising of the $\mathcal{N}=2$
central charge function $Z$: non-BPS $Z\neq 0$ and non-BPS $Z=0$ (see
\textit{e.g.} \cite{BFGM1} for a detailed analysis in the case of symmetric
SK geometries).

When considering $\mathcal{N}=2$, $d=4$ supergravity coming from a string
compactification, the perturbative quantum corrections to the holomorphic
prepotential of SK geometries can be polynomials or some non-polynomial
(usually polylogarithmic) functions of the moduli (see \textit{e.g.} \cite
{Quantum-N=2} and Refs. therein). In the case of cubic classical SK
geometries, such as the ones obtained in the large volume limit of $CY_{3}$%
-compactifications of Type IIA, the sub-leading nature of the quantum
corrections constrains the most general polynomial correction to be at most
of degree two in the moduli, with a priori complex coefficients.

In \cite{CFG} it has been shown that the only polynomial quantum
perturbative correction to the prepotential of classical cubic SK geometries
which is consistent with the perturbative (continuous) axion-shift symmetry
\cite{Peccei-Quinn} is the constant purely imaginary term ($i=1,...,n_{V}$
throughout):
\begin{equation}
\mathcal{F}_{classical}=d_{ijk}t^{i}t^{j}t^{k}\longrightarrow \mathcal{F}%
_{quantum-pert.}=d_{ijk}t^{i}t^{j}t^{k}+i\lambda ,~\lambda \in \mathbb{R},
\end{equation}
where $d_{ijk}$ is the real, constant, completely symmetric tensor defining
the cubic geometry. Indeed, it can be easily shown that all other polynomial
perturbative corrections (quadratic, linear and real constant terms in the
moduli) do not modify the classical cubic geometry, since they do not
contribute to the K\"{a}hler potential at all \cite{CFG}.

Let us start by considering the holomorphic prepotential of a certain class
of compactifications of the heterotic $E_{8}\times E_{8}$ superstring over $%
K_{3}\times T^{2}$, having the form (see \textit{e.g.} \cite{Quantum-N=2}
and refs. therein)
\begin{equation}
\mathcal{F}^{heterotic}=stu-s\sum_{a=4}^{n_{V}}\left( \widetilde{t}%
^{a}\right) ^{2}+h_{1-loop}\left( t,u,\widetilde{t}\right)
+f_{non-pert.}\left( e^{-2\pi s},t,u,\widetilde{t}\right) .
\label{F-heterotic}
\end{equation}
Since in this case the dilaton $s$ belong to a vector multiplet, there exist
($T$-symmetric) quantum perturbative string-loop corrections and
non-perturbative corrections, as well. The tree-level, classical term $%
stu-s\sum_{a=4}^{n_{V}}\left( \widetilde{t}^{a}\right) ^{2}$ is the
prepotential of the generic cubic sequence $\frac{SU(1,1)}{U\left( 1\right) }%
\otimes \frac{SO\left( 2+n,2\right) }{SO\left( 2+n\right) \otimes SO\left(
2\right) }$ ($n_{V}=n+3$) of homogeneous symmetric SK manifolds (see \textit{%
e.g.} \cite{CFG}, and \cite{BFGM1} and Refs. therein), in the symplectic
gauge exhibiting the largest possible amount of explicit symmetry $SO\left(
n+1,1\right) $. Due to non-renormalization theorems, all the quantum
perturbative string-loop corrections are encoded in the 1-loop contribution $%
h_{1-loop}$, which contains a constant term, a purely cubic polynomial term
and a polylogarithmic part (see \textit{e.g.} \cite{Quantum-N=2} and refs.
therein). Finally, $f_{non-pert.}$ encodes the non-perturbative corrections,
exponentially suppressed in the limit $s\rightarrow \infty $.

By exploiting the Type IIA/heterotic duality and correspondingly by suitably
identifying the relevant moduli fields, the heterotic prepotential (\ref
{F-heterotic}) becomes structurally identical to the one arising from Type
IIA compactifications over Calabi-Yau threefolds ($CY_{3}$s). In such a
case, the prepotential of the resulting low-energy $\mathcal{N}=2$, $d=4$
supergravity is purely of classical origin. Indeed, there are only
K\"{a}hler structure moduli, and the dilaton $s$ belongs to an
hypermultiplet; thus, there are no string-loop corrections, and all
corrections to the large-volume limit cubic prepotential come from the
world-sheet sigma-model \cite{Alvarez-Gaume}. As shown in \cite{Grisaru,
CDLOGP1,CDLOGP2}\textbf{,} there are no 1-, 2- and 3-loop contributions.
Moreover, the non-perturbative, world-sheet instanton corrections destroy
continuous axion-shift symmetry, by making it discrete \cite{Peccei-Quinn}.

By disregarding such non-perturbative world-sheet instanton corrections, the
Type IIA prepotential can be written as follows ($n_{V}=h_{1,1}$) \cite
{CDLOGP1,CDLOGP2,HKTY,Quantum-N=2}:
\begin{equation}
\mathcal{F}^{IIA}=\frac{1}{3!}\mathcal{C}_{ijk}t^{i}t^{j}t^{k}+\mathcal{W}%
_{0i}t^{i}-i\frac{\chi \zeta \left( 3\right) }{16\pi ^{3}}.
\end{equation}
The $\mathcal{C}_{ijk}$ are the real classical intersection numbers,
determining the large volume limit cubic SK geometry. The perturbative
contributions from $2$-dimensional CFT on the world-sheet are encoded in a
linear and in a constant term.

The linear term is determined by the $\mathcal{W}_{0i}=\frac{1}{4!}%
c_{2}\cdot J_{i}=\frac{1}{4!}\int_{CY_{3}}c_{2}\wedge J_{i}$, which are the
real expansion coefficients of the second Chern class $c_{2}$ of $CY_{3}$
with respect to the basis $J_{i}^{\ast }$ of the cohomology group $%
H^{4}\left( CY_{3},\mathbb{R}\right) $, dual to the basis of the (1,1)-forms
$J_{i}$ of the cohomology $H^{2}\left( CY_{3},\mathbb{R}\right) $. It has
been shown that the linear term $\mathcal{W}_{0i}t^{i}$ can be reabsorbed by
a suitable symplectic transformation of the period vector, and thus in the
heterotic picture it has just the effect of a constant shift in $\Im s$ (
\cite{Witten-theta}\textbf{; }see also \cite{Quantum-N=2}).

From the general analysis of \cite{CFG}, the constant term is the only
relevant one. It is determined by the Euler character\footnote{%
It is worth remarking that for typical $CY_{3}$s $\left| \chi \right|
\leqslant 10^{3}$, and thus $\frac{\chi \zeta \left( 3\right) }{16\pi ^{3}}$
is of order $1$. Notice that $\chi =0$ for \textit{self-mirror} $CY_{3}$s,
and thus such a constant term vanishes. Moreover, for some particular
\textit{self-mirror} models, such as the so-called FHSV one \cite{FHSV},
also the non-perturbative, world-sheet instanton corrections vanish; thus,
in such models, up to suitable symplectic transformations of the period
vector, the classical cubic prepotential does not receive any perturbative
and non-perturbative correction.} $\chi $ of $CY_{3}$ ($\zeta $ is the
Riemann zeta-function), and it has a 4-loop origin in the non-linear
sigma-model \cite{Alvarez-Gaume, Grisaru, CDLOGP1,CDLOGP2}.

It is worth pointing out here that $CY_{3}$-compactifications of
Type IIB do not admit a large volume limit; moreover, the Attractor
Eqs. only depend on the complex structure moduli (which are the
scalars of the $\mathcal{N}=2$ vector multiplets). The solutions to
$\mathcal{N}=2$, $d=4$ Attractor Eqs. for the resulting SK
geometries were studied in \cite{BFMY} for the particular class of Fermat $%
CY_{3}$s with $n_{V}=1$, and in \cite{Misra1} for a particular $CY_{3}$ with
$n_{V}=2$.

Aim of the present paper is to study the solutions to the $\mathcal{N}=2$, $%
d=4$ Attractor Eqs. in a dyonic background in the simplest case of
perturbative quantum corrected cubic SK geometry, namely in the $1$-modulus
SK geometry described (in a suitable special symplectic coordinate) by the
holomorphic K\"{a}hler gauge-invariant prepotential $\mathcal{F}%
=t^{3}+i\lambda $, with $\lambda \in \mathbb{R}$ (see Eq. (\ref{t^3+ic})
below). By doing so, we will extend the BPS analysis of \cite{Quantum-N=2}.

The plan of the paper is as follows.

In Sect. \ref{Classical} we shortly review the classical so-called $t^{3}$
model, focussing on the attractors supported by the ``magnetic'' BH charge
configuration, in which the charge vector $Q\equiv \left(
q_{0},q_{1},p^{0},p^{1}\right) =\left( q_{0},0,0,p^{1}\right) $. Thence, in
Sect. \ref{Quantum} we will study the solutions to $\mathcal{N}=2$, $d=4$
Attractor Eqs. for the $t^{3}+i\lambda $ model, dividing our analysis in the
classical BPS (Subsect. \ref{Quantum-BPS}) and non-BPS (Subsect. \ref
{Quantum-non-BPS}) charge domains. Despite the simplicity of the correction
added to the classical prepotential $\mathcal{F}=t^{3}$ (see Eq. (\ref{t^3})
below), we will find evidence for interesting phenomena, such as the \textit{%
``separation''} and the \textit{``transmutation'' of extremal BH
attractors}. Finally, in Subsect. \ref{D0-D6} we briefly consider
the so-called $D0-D6$ BH charge configuration $Q=\left(
q_{0},0,p^{0},0\right) $, pointing out that it does not support
admissible BPS attractors, both at the classical and quantum level.
An Appendix, containing some technical details, concludes the paper.

\section{\label{Classical}$t^{3}$ model}

The classical so-called $t^{3}$ model is a $1$-modulus model based on a
cubic holomorphic prepotential which (in the local special symplectic
coordinate $t$) reads
\begin{equation}
\mathcal{F}(t)=t^{3},  \label{t^3}
\end{equation}
constrained by the condition
\begin{equation}
\Im t<0.  \label{classKCons}
\end{equation}
The corresponding manifold is the rank-$1$ symmetric special K\"{a}hler (SK)
space $\frac{SU\left( 1,1\right) }{U\left( 1\right) }$. This is an isolated
case in the classification of the symmetric SK manifolds (see \textit{e.g.}
\cite{CFG}, and \cite{BFGM1} and Refs. therein). Furthermore, it is worth
pointing out that $\frac{SU\left( 1,1\right) }{U\left( 1\right) }$ can be
endowed with a K\"{a}hler gauge-invariant quadratic prepotential $\mathcal{F}%
=\frac{i}{4}\left( t^{2}-1\right) $, as well; in such a case, it corresponds
to the $n=0$ element of the sequence of symmetric irreducible quadratic SK
manifolds $\frac{SU(1,1+n)}{U(1)\otimes SU(1+n)}$ ($n=n_{V}-1$, see \textit{%
e.g.} \cite{CFG}, and \cite{BFGM1} and Refs. therein), with geometric
properties completely different from the cubic case (see \textit{e.g.} the
discussion in \cite{BFM-SIGRAV06}). The reason for such a difference lies in
the fact that the four charges $q_{0},q_{1},p^{0},p^{1}$ sit in the spin $%
\frac{3}{2}$ real representation of $SU\left( 1,1\right) $ in the former
case, and in the spin $\frac{1}{2}$ complex representation of $SU\left(
1,1\right) $ in the latter case.

At the present time, the $t^{3}$ model is the only model whose $\mathcal{N}%
=2 $, $d=4$ Attractor Eqs. have been solved for a completely generic BH
charge configuration $\left( q_{0},q_{1},p^{0},p^{1}\right) $. Its $\frac{1}{%
2}$-BPS solutions were known after \cite{Shmakova}, whereas in \cite{TT}\
its non-BPS solutions with non-vanishing $\mathcal{N}=2$ central charge $Z$
were determined for $q_{1}=0$. Recently, in \cite{Saraikin-Vafa-1} such a
result was extended to the general case $q_{1}\neq 0$.

Since such a model does not admit non-BPS $Z=0$ attractors, in the following
``non-BPS'' will be understood for ``non-BPS $Z\neq 0$''. Furthermore, we
will always consider the simple case corresponding to the so-called
``magnetic'' charge configuration, in which the only non-vanishing BH
charges are $q_{0}$ and $p^{1}$ (in a stringy interpretation, this
corresponds to consider only $D0$ and~$D4$ branes).

In the remaining of this Section we will report the main known results about
$t^{3}$ model in ``magnetic'' BH charge configuration, in order to make the
comparison with the quantum-perturbed case easier.

The K\"{a}hler potential and metric function are given by
\begin{equation}
K=-ln\left[ -8(\Im t)^{3}\right] ,\qquad g\equiv \partial _{t}\overline{%
\partial }_{\overline{t}}K=\frac{3}{4(\Im t)^{2}};  \label{FKGclass}
\end{equation}
thus the condition (\ref{classKCons}) is nothing but the reality condition
for $K$. Notice that $g>0$ for $\Im t\neq 0$, and in particular in the lower
half of the Argand-Gauss plane $\mathbb{C}$ determined by (\ref{classKCons}%
). The superpotential, its covariant derivative and the effective black hole
(BH) potential are respectively given by
\begin{equation}
\begin{array}{l}
\displaystyle W=q_{0}-3p^{1}t^{2},\qquad D_{t}W=-6p^{1}t+\frac{3i}{2\Im t}W;
\\
\displaystyle V_{BH}=-\frac{3\left( \left( \Im t\right) ^{2}p^{1}\right)
^{2}+12\left( p^{1}\Im t\,\Re t\right) ^{2}+\left( q_{0}-3p^{1}\left( \Re
t\right) ^{2}\right) ^{2}}{2\Im t^{3}}.
\end{array}
\end{equation}
The BPS and non-BPS solutions to the Attractor Eq.
\begin{equation}
\frac{\partial V_{BH}}{\partial t}=0  \label{AE}
\end{equation}
are always stable, and they are supported by BH charges satisfying $%
p^{1}q_{0}>0$ and $p^{1}q_{0}<0$, respectively\footnote{%
It is easy to recognize $p^{1}q_{0}>0$ and $p^{1}q_{0}<0$ respectively as
the ``magnetic'' branches of the 2 homogeneous symmetric BH charge BPS and
non-BPS $Z\neq 0$ orbits of the symplectic vector representation space of
the $U$-duality group $SU\left( 1,1\right) $ of the $t^{3}$ model (see
Appendix II of \cite{BFGM1}).} \cite{Shmakova,TT,Saraikin-Vafa-1}. They are
both axion-free:
\begin{equation}
\left. \Re t\right| _{cr.}=0,\qquad \left. \Im t\right| _{cr.}=\sqrt{\,%
\rule[-0.75em]{0.2pt}{2em}\frac{q_{0}}{p^{1}}\rule[-0.75em]{0.2pt}{2em}\,},
\label{classical-solutions}
\end{equation}
and the corresponding BH entropy reads
\begin{equation}
S_{BH}=\pi V_{BH,cr.}=2\pi \sqrt{\rule[-0.4em]{0.2pt}{1.4em}\,q_{0}\left(
p^{1}\right) ^{3}\rule[-0.4em]{0.2pt}{1.4em}\,}.  \label{classical-SBH}
\end{equation}

\section{\label{Quantum}$t^{3}+i\protect\lambda $ Model}

From the considerations made in the Introduction, it follows that the most
general form of $1$-modulus cubic geometry with polynomial quantum
perturbative corrections consistent with axion-shift symmetry is based on
the holomorphic K\"{a}hler gauge invariant prepotential function
\begin{equation}
\mathcal{F}(t)=t^{3}+i\lambda ,~\lambda \in \mathbb{R}.  \label{t^3+ic}
\end{equation}

Let us start by noticing that the SK manifold based on the prepotential (\ref
{t^3+ic}) is no more symmetric nor homogeneous. Its K\"{a}hler potential and
metric function have the following form:
\begin{equation}
K=-ln\left[ -4\lambda -8\left( \Im t\right) ^{3}\right] ,\qquad g=3\frac{%
\left( \strut (\Im t)^{3}-\lambda \right) \Im t}{\left( \lambda +2(\Im
t)^{3}\right) ^{2}}.  \label{geom-t^3+ic}
\end{equation}
In the ``magnetic'' BH charge configuration, the superpotential, its
covariant derivative and the effective BH potential respectively reads
\begin{equation}
W=q_{0}-3p^{1}t^{2},\quad D_{t}W=-6p^{1}t+\frac{3i\left( \Im t\right) ^{2}W}{%
\lambda +2\left( \Im t\right) ^{3}},\qquad V_{BH}=e^{K}\left[ W\bar{W}%
+g^{-1}D_{t}W\overline{\strut D_{t}W}\,\right] .  \label{WK}
\end{equation}
When switching $\lambda \neq 0$, the reality condition~(\ref{classKCons}) on
the K\"{a}hler potential gets modified as follows:
\begin{equation}
\Im t<-\sqrt[3]{\frac{\lambda }{2}}.
\end{equation}
Furthermore, the positivity of the metric function yields another
restriction on allowed values of $\Im t$:
\begin{equation}
\qquad \left( \strut (\Im t)^{3}-\lambda \right) \Im t>0.
\end{equation}
\begin{figure}[h]
$\epsfxsize=0.45\textwidth\epsfbox{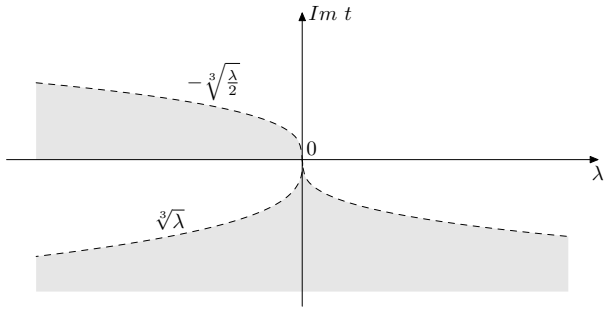}$%
\caption{Admissible regions for $\Im t$}
\label{regions}
\end{figure}

Thus, $\Im t$ belong to the shaded regions represented in Fig.~\ref{regions}%
. If~$\lambda >0$, the allowed region consists of a single interval,
otherwise it is the union of two non-overlapping intervals:
\begin{equation*}
\begin{array}{ll}
\displaystyle\lambda >0: & \qquad \Im y\in \left( -\infty ,-\sqrt[3]{\frac{%
\lambda }{2}}\;\right) \\
\displaystyle\lambda <0: & \qquad \Im y\in \left( -\infty ,\sqrt[3]{\strut
\lambda }\;\right) \,\bigcup \,\left( 0,-\sqrt[3]{\frac{\lambda }{2}}%
\;\right) .
\end{array}
\end{equation*}

It is instructive to consider the $\lambda $-dependence of the Ricci scalar
curvature $R$. Whereas in the classical case $R=-\frac{2}{3}$, in the
considered quantum case $R$ is not constant nor necessarily negative
anymore:
\begin{equation}
R=\frac{\lambda ^{4}+32\lambda ^{3}\left( \Im t\right) ^{3}-48\lambda
^{2}\left( \Im t\right) ^{6}+104\lambda \left( \Im t\right) ^{9}-8\left( \Im
t\right) ^{12}}{12\left( \Im t\right) ^{3}(-\lambda +\left( \Im t\right)
^{3})^{3}}.
\end{equation}
For $\lambda >0$ the $\Im t\rightarrow \infty $ limit gives~$R=-2/3$; on the
other hand, $R$ decreases approaching the boundary value~$\Im t=-\sqrt[3]{%
\lambda /2}$, where it equals $-2$.

For $\lambda <0$, $R$ increases starting from its value at $\Im t\rightarrow
-\infty $, and it diverges in the limit $\Im t\rightarrow \sqrt[3]{\strut
\lambda }^{-}$. In the other interval $\left( 0,-\sqrt[3]{\frac{\lambda }{2}}%
\;\right) $, $R$ diverges in the limit $\Im t\rightarrow 0^{+}$, whereas it
reaches its minimal value~$-2$ at the boundary $\Im t=-\sqrt[3]{\frac{%
\lambda }{2}}$.

Concerning the effective BH potential, it is a real function of one complex
variable $t$, and it contains three parameters: the charges $q_{1}$ and $%
p^{0}$ and the quantum parameter $\lambda $. It enjoys the following
properties with respect to reflection of its arguments:
\begin{equation}
V_{BH}(-t;p^{1},q_{0},-\lambda )=-V_{BH}(t;p^{1},q_{0},\lambda ),\qquad
V_{BH}(t;-p^{1},-q_{0},\lambda )=V_{BH}(t;p^{1},q_{0},\lambda ).
\label{VBHreflection}
\end{equation}

In order to decrease the number of independent parameters and simplify the
analysis, in the following treatment we will redefine $t$ and $\lambda $ in
such a way to factorize the dependence of $V_{BH}$ on the charges.

The second reflection property shows that $V_{BH}$ is somehow sensitive to
the sign of~$p^{1}q_{0}$. This fact, in light of the form of the
``magnetic'' supporting BH charge orbits in the classical case, leads us to
divide the treatment of the quantum model in two parts, respectively
corresponding to:

\begin{enumerate}
\item  $p^{1}q_{0}>0$ and refer to this range of charges as~\textit{%
classical BPS charge domain} (or shortly \textit{BPS domain}; for $\lambda
=0 $ it supports only BPS attractors);

\item  $p^{1}q_{0}<0$ and refer to this range of charges as~\textit{%
classical non-BPS charge domain }(or shortly \textit{non-BPS domain}; for $%
\lambda =0$ it supports only non-BPS attractors).
\end{enumerate}

\subsection{\label{Quantum-BPS}BPS domain}

Let us switch to a new coordinate~$y$ and to a new quantum parameter~$\alpha
$:
\begin{equation}
t\equiv yp^{1}\sqrt{\frac{q_{0}}{(p^{1})^{3}}},\qquad \lambda \equiv \alpha
q_{0}\sqrt{\frac{q_{0}}{(p^{1})^{3}}},  \label{def-BPS}
\end{equation}
in terms of which the dependence of all the considered quantities on the
charges is factorized\footnote{%
For brevity's sake, we write~$\Im y^{n}$ instead of~$(\Im y)^{n}$. The same
holds for~$\Re y$.}:
\begin{equation}
\begin{array}{l}
\displaystyle W=q_{0}\left( 1-3y^{2}\right) ,\qquad e^{-K}=-4q_{0}\sqrt{%
\frac{q_{0}}{(p^{1})^{3}}}\left( \alpha +2\Im y^{3}\right) ; \\[0.5em]
\displaystyle g=\frac{3p^{1}}{q_{0}}\,\frac{\left( (\Im y)^{3}-\alpha
\right) \Im y}{\left( 2(\Im y)^{3}+\alpha \right) ^{2}},\qquad V_{BH}=v(y,%
\bar{y},\alpha )\frac{q_{0}}{\displaystyle\sqrt{\frac{q_{0}}{(p^{1})^{3}}}};
\\[0.5em]
\displaystyle v(y,\bar{y},\alpha )\equiv \frac{1}{4\Im y\left( \alpha
^{2}+\alpha \Im y^{3}-2\Im y^{6}\right) }\left[ \rule[-1em]{0pt}{2.5em}%
12\alpha ^{2}\left( \Im y^{2}+\Re y^{2}\right) \right. \\[0.5em]
\displaystyle\phantom{V_{eff}=4\Im y\left( \alpha^2 \alpha \Im y^3 \right)}%
\left. +4\Im y^{4}\left( 3\Im y^{4}+12\Im y^{2}\Re y^{2}+\left( 1-3\Re
y^{2}\right) ^{2}\right) \right. \\
\displaystyle\phantom{V_{eff}=4\Im y\left( \alpha^2 \alpha \Im y^3 \right)}%
\left. +\alpha \Im y\left( 3\Im y^{4}-\left( 1-3\Re y^{2}\right) ^{2}-6\Im
y^{2}\left( 3+\Re y^{2}\right) \right) \rule[-1em]{0pt}{2.5em}\right] .
\end{array}
\label{WKgBPSD}
\end{equation}
As stated above, the topology of the allowed regions of $\Im t$ depends on
the sign of $\lambda $, as given by Fig.~\ref{regions}. In terms of $y$ and $%
\alpha $ defined in the BPS domain by Eq. (\ref{def-BPS}), the corresponding
allowed regions are represented in Fig.~\ref{BPSregions}.

\begin{figure}[h]
$\epsfxsize=0.45\textwidth\epsfbox{allowedRegionBPS.1}$ \quad $\epsfxsize%
=0.45\textwidth\epsfbox{allowedRegionBPS.2}$%
\caption{Domains of positivity of $g$ and $e^{K}$}
\label{BPSregions}
\end{figure}
For example, if one considers~$q_{0}<0$ then for $\alpha >0$ there are two
disconnected allowed intervals for $\Im y$:
\begin{equation}
\Im y\in \left( -\infty ,\sqrt[3]{\strut \alpha }\;\right) \,\bigcup
\,\left( 0,-\sqrt[3]{\frac{\alpha }{2}}\;\right) ,
\end{equation}
while for $\alpha <0$ it can range in only one interval:
\begin{equation}
\Im y\in (-\sqrt[3]{\strut \frac{\alpha }{2}},\infty ).
\end{equation}

In the classical case there is a one-to-one correspondence between the
symplectic vector of BH charges inserted as input in the Attractor Eq. (\ref
{AE}) and the solutions to this equation. Thus, for example in the BPS
domain $p^{0}q_{1}>0$, when inserting two arbitrary values for $p^{0}$ and $%
q_{1}$ of the same sign, in the classical case there exists one and only one
solution (in particular, of BPS type) to the classical Attractor Eq. (\ref
{AE}).

This is no more true in the considered quantum case: in the BPS domain $%
p^{0}q_{1}>0$, when inserting two arbitrary values for $p^{0}$ and $q_{1}$
of the same sign, now there exist two stable critical points\footnote{%
Also unstable non-BPS critical points of $V_{BH}$ may exist; however, we
will not deal with them, because they do not determine \textit{attractors }%
in strict sense.} of $V_{BH}$, one BPS and the other non-BPS. It is worth
remarking that in the considered BPS domain the non-BPS critical points of $%
V_{BH}$ do not admit a classical limit $\alpha \rightarrow 0$.

Furthermore, while in the classical case both BPS and non-BPS critical
points of $V_{BH}$ supported by a ``magnetic'' BH charge configuration are
axion-free, in the quantum case also non-BPS critical points of $V_{BH}$
with non-vanishing axion may arise out (and they can also be stable). As in
the classical case, also in the considered quantum case non-BPS critical
points of $V_{BH}$ with $Z=0$ do not exist at all. A pictorial view of the
more complicated situation typical of the quantum case is given by Fig.~\ref
{BPSSolutions}. As mentioned in the Appendix, in both domains (BPS and
non-BPS) of the quantum case the BPS solutions are known analytically,
whereas the non-BPS solutions can be investigated only numerically.
\begin{figure}[h]
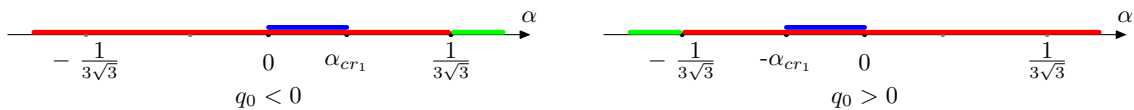

\begin{tabular}{lcr}
$\epsfxsize=0.4\textwidth\epsfbox{solutions.1}$ & \qquad & $\epsfxsize=0.4%
\textwidth\epsfbox{solutions.2}$%
\end{tabular}
\caption{Ranges of the quantum parameter $\protect\alpha $ supporting minima
of $V_{BH}$, for the cases $q_{0}<0$ and $q_{0}>0$. The red line corresponds
to BPS minima, the green one to axion-free non-BPS minima, and the blue one
to non-BPS minima with non-vanishing axion}
\label{BPSSolutions}
\end{figure}
\newline

Axion-free stable non-BPS critical points arise only for $\alpha >\frac{1}{3%
\sqrt{3}}$ (in the case $q_{0}<0$) and for~$\alpha <-\frac{1}{3\sqrt{3}}$
(in the case $q_{0}>0$). On the other hand, non-BPS critical points with
non-vanishing axion field exists only for $0<\alpha <\alpha _{cr_{1}}$ (in
the case $q_{0}<0$) and for~$-\alpha _{cr_{1}}<\alpha <0$ (in the case $%
q_{0}>0$), where the value of the $\alpha _{cr_{1}}$ is known just
numerically:~$\alpha _{cr_{1}}\in \left( 0.030101,0.030102\right) $.

By looking at Fig.~\ref{BPSSolutions}, one can realize at a glance
that two new phenomena, absent in the classical case, arise in the
quantum case. The first one, which we will name
\textit{``separation'' of attractors}, occurs when the value of the
quantum parameter is fixed in a certain range. The second one, which
we will call \textit{``transmutation'' of attractors}, occurs when
the value of the quantum parameter is varied, and becomes greater
(or smaller) of a certain critical value.

Concerning the \textit{``separation''}, from Fig.~\ref{BPSSolutions}
one sees that for~$|\alpha |<\alpha _{cr_{1}}$ two different stable
critical points of $V_{BH}$ exist for the same BH charge
configuration: one of them is BPS (whose classical limit is the well
known BPS solution (\ref {classical-solutions})), the other one
is~non-BPS with~$\Re y\neq 0$ (which does not have a classical
limit). Such a \textit{``separation''} of the solutions of the
Attractor Eq. (\ref{AE}) can ultimately be traced back to the
presence of two disconnected regions in the domain of allowed values
for $\Im y$ (see above treatment), which determines two disconnected
regions of asymptotical ($r\rightarrow \infty $) values for $\Im y$
(also called \textit{``area codes''} or \textit{``basins of
attraction''} in literature \cite{Kal1, Kal2,Moore,G,Misra2}). The
dynamical radial evolution of the modulus $t$ is completely
deterministic, and determines one unique solution for an arbitrary
but fixed ``magnetic'' BH charge configuration given as input to the
Attractor Eq. (\ref{AE}), \textit{but only inside each ``area
code''}. Thus, the core of the Attractor Mechanism is preserved in
the considered framework of dyonic, extremal, static, spherically
symmetric and asymptotically flat BHs in $\mathcal{N}=2$, $d=4$
supergravity. In other words, the Attractor Mechanism gets two-fold
split due to the presence of two disconnected regions in the allowed
values of the modulus $t$, separated by a region where the metric
function $g$ is negative. Such a result allows one to conjecturally
argue that in presence of $m$ disconnected regions allowed in the
moduli space, in general there exist $m$ solutions to the
Attractor Eqs. for an arbitrary but fixed supporting BH charge configuration%
\footnote{%
In the context of $n_{V}=1$ SK geometries, the phenomenon of \textit{%
``separation''} of the solutions of the Attractor Eq. (\ref{AE}) can
be observed also for more general geometries, which do not
corrispond to cubic geometries in the large modulus limit.
\par
An example is given by the $1$-modulus SK geometry based on the holomorphic
prepotential $\mathcal{F}=t^{4}$ (we thank Mario Trigiante for discussions
on this issue).}. A similar phenomenon was observed some years ago in $%
\mathcal{N}=2$, $d=5$ supergravity in \cite{Kal1,Kal2}; the disconnectedness
of the moduli space preserves the validity of the Attractor Mechanism inside
\textit{each} allowed connected region, and therefore the results of
existence and uniqueness of the solutions to the Attractor Eqs. obtained in
\cite{Zhukov} are still valid, but inside \textit{each} \textit{``area code''%
}.

Concerning the \textit{``transmutation''}, let us take~$q_{0}<0$ (this does
not imply any loss of generality). Thence, from Fig.~\ref{BPSSolutions} one
sees that varying the quantum parameter $\alpha $ across the critical value $%
\frac{1}{3\sqrt{3}}$ the BPS stable critical point \textit{``transmutes''}
into the non-BPS critical point (or \textit{vice versa}).

By recalling that $\alpha $ depends on (a dimensionless ratio of) $q_{0}$
and $p^{1}$, it is easy to realize that the occurrence of such a phenomenon
can be traced back to the fact that the supporting ``magnetic'' (branches of
the) BH charge orbits still exist in the considered quantum case, but, as
the quantum moduli space itself, they are not symmetric nor homogeneous
manifolds any more.

As mentioned above, since the BPS critical points can be computed
analytically, one can calculate the $\alpha $-dependent expression of the
BPS BH entropy in the BPS domain to be
\begin{equation}
S_{BH}=\pm \frac{\pi }{4}\frac{\left( 1+3\Im y^{2}\right) ^{2}}{\alpha +2\Im
y^{3}},\quad \mbox{with $\Im y$ satisfying}\quad \Im y^{3}-\Im y+2\alpha =0,
\end{equation}
where the solution of the cubic Eq. must be chosen inside the allowed
region(s) of the moduli space, and the $\pm $ branches of $S_{BH}$ must be
chosen in order to obtain $S_{BH}>0$. In the classical limit $\alpha
\rightarrow 0$, one recovers the well known value~$S_{BH}=2\pi \sqrt{\strut
q^{0}(p_{1})^{3}}$ given by Eq. (\ref{classical-SBH}).

Let us now analyze the evolution $V_{BH}$ with respect to the quantum
parameter $\alpha $. Choosing $q_{0}<0$ without loss of generality, we
consider $\alpha >0$, because, due to the presence of the \textit{%
``separation'' of attractors} described above, it is the most
interesting case.

\begin{figure}[h]
\includegraphics[width=0.45\textwidth]{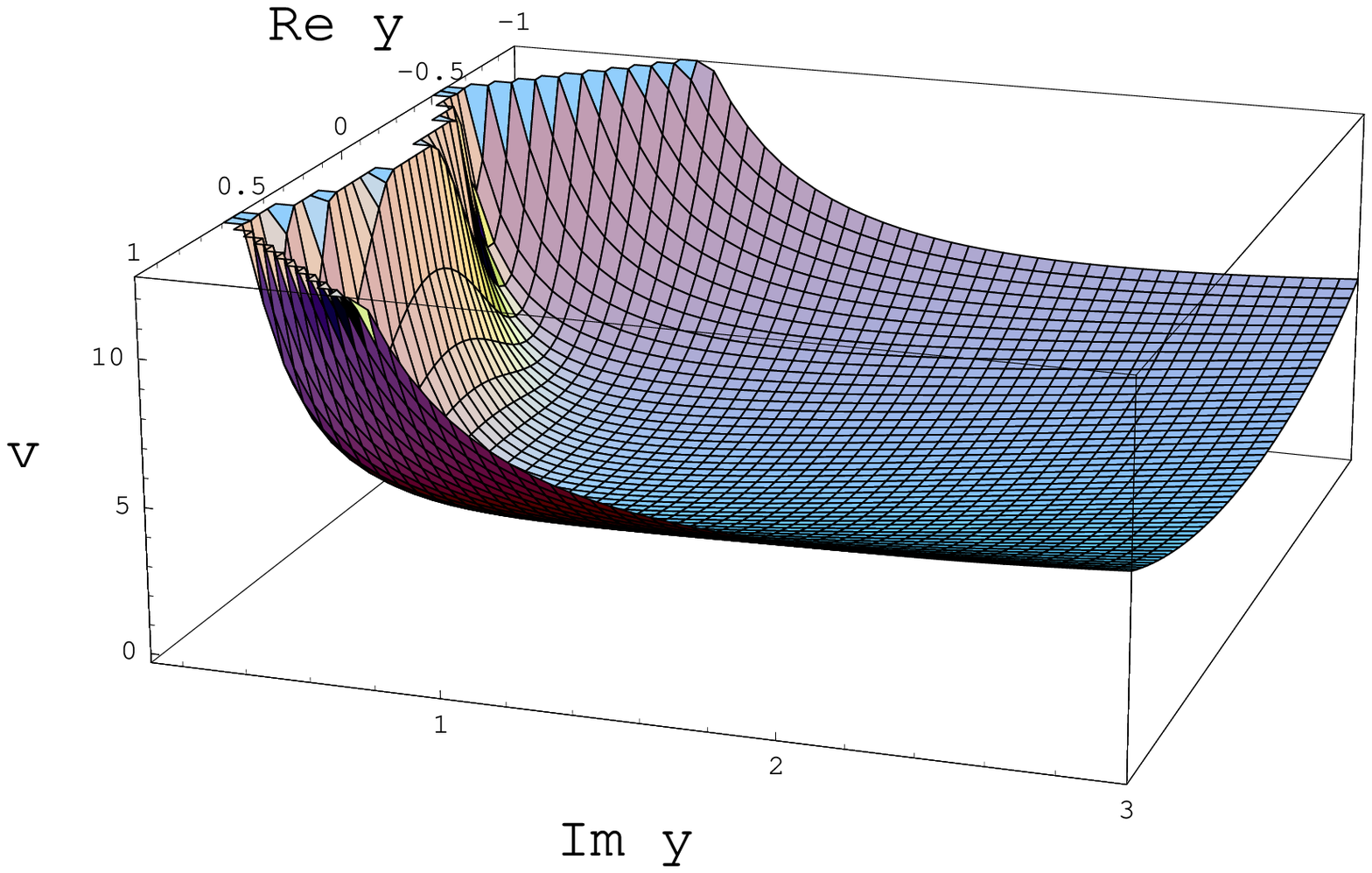}
\includegraphics[width=0.45\textwidth]{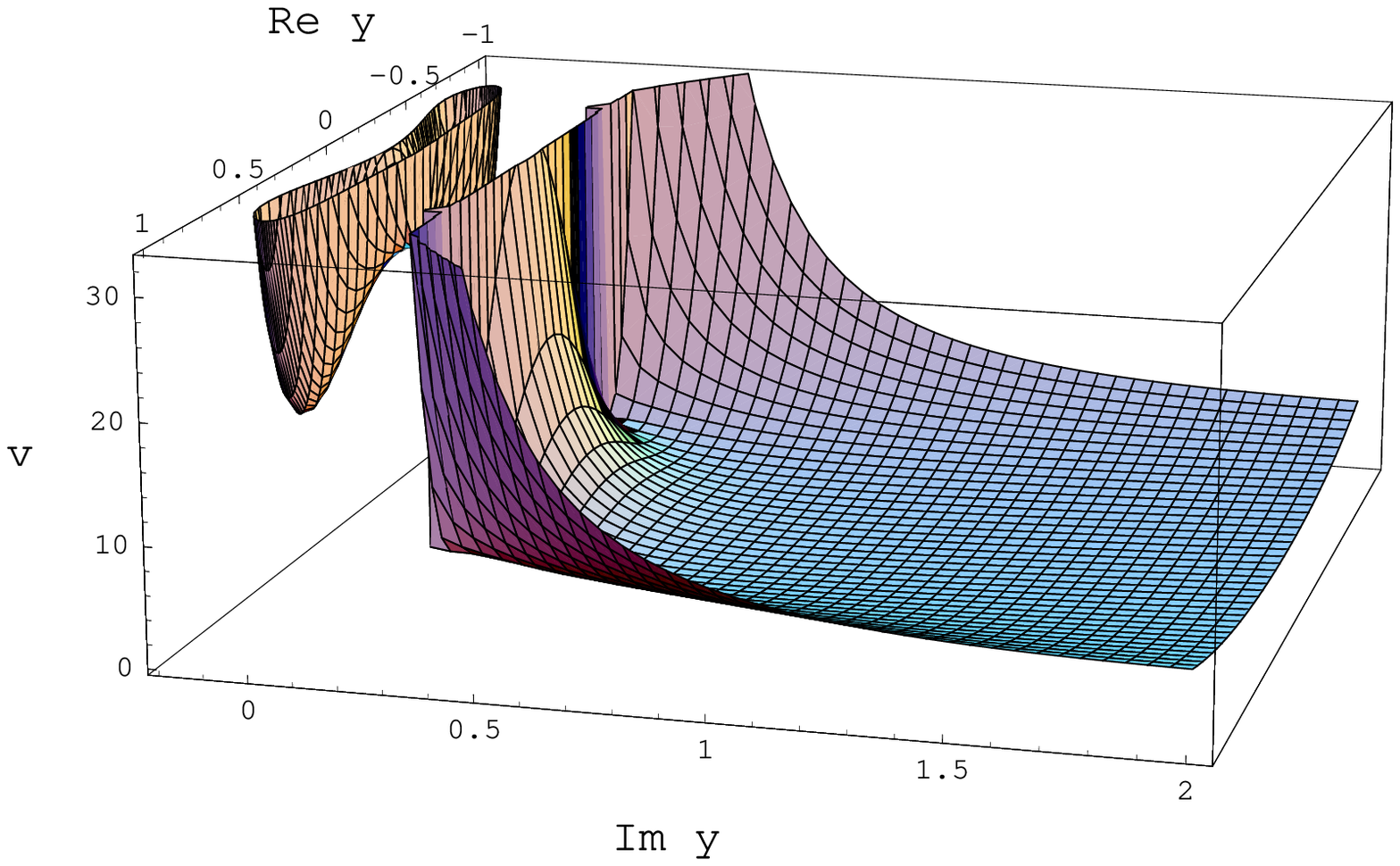}
\caption{Plots of $V_{BH}$ for the classical case~($\protect\alpha =0$,
left) and~for $\protect\alpha =1/100$ (right) }
\label{VBHBPSPlot}
\end{figure}

The classical limit of the function~$v$ defined in Eq. (\ref{WKgBPSD}) has
the form of a ``scoop'', with a minimum at the point~$\Re y=0$, $\Im y=1$
(see Eq. (\ref{classical-solutions}) and Fig.~\ref{VBHBPSPlot}).

Let us now switch the (considered class of) quantum corrections on,
and slightly increase the value of $\alpha $ from $0$ to $0.01$. As
one can see from Fig.~\ref{BPSSolutions}, $\alpha =0.01$ is in the
range supporting the \textit{``separation'' of attractors},\textit{\
i.e. }the coexistence of the axion-free BPS attractors with the
non-BPS ones with non-vanishing axion. Such stable critical points
of $V_{BH}$ respectively have coordinates
\begin{equation}
\begin{array}{l}
BPS:\quad \Re y=0,\quad \Im y\approx 0.99,\qquad V_{BH,BPS}\approx 1.99\sqrt{%
\strut q_{0}(p^{1})^{3}}; \\
\\
non-BPS:\quad \Re y\approx \pm 0.51,\quad \Im y\approx -0.08,\qquad
V_{BH,non-BPS}\approx 15.54\sqrt{\strut q_{0}(p^{1})^{3}}.
\end{array}
\label{comparison-1}
\end{equation}

As one can see from the plot of $V_{BH}$, in contrast to the classical case
there appears a new chump of the form of camel humps, which contains a
stable non-BPS critical point (not existing in the classical limit). The
width of such a disconnected branch of $V_{BH}$ is~$\sqrt[3]{\alpha /2}$,
and thus it vanishes in the classical limit. The region of separation
between the two branches of $V_{BH}$ has width~$\sqrt[3]{\alpha }$, and in
such a region $g<0$.

As yielded by Eq. (\ref{comparison-1}), for $\alpha =0.01$ (and
actually in the whole range of $\alpha $ supporting the
\textit{``separation'' of attractors}) it holds that
$V_{BH,BPS}<V_{BH,non-BPS}$. It is worth remarking that this is the
opposite of what happens in globally supersymmetric field theories
with multiple local minima, where the non-supersymmetric stable
critical points of the superpotential are energetically favored with
respect to their supersymmetry-preserving counterparts (see
\textit{e.g.} \cite{Seiberg,Ooguri-Vafa,Ooguri-Erice} and Refs.
therein).

\begin{figure}[h]
\begin{tabular}{cc}
\includegraphics[width=0.5\textwidth]{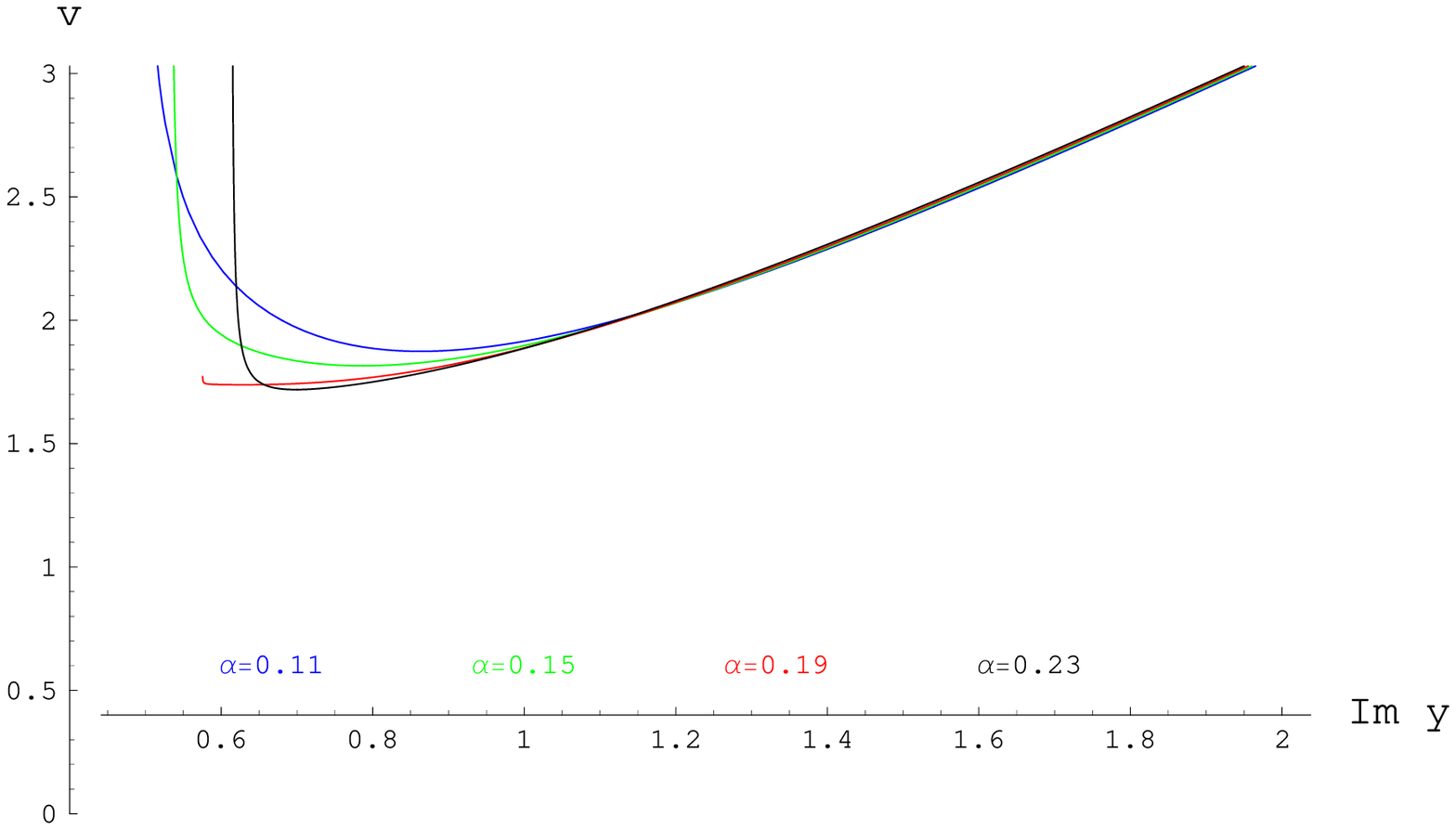} & %
\includegraphics[width=0.5\textwidth]{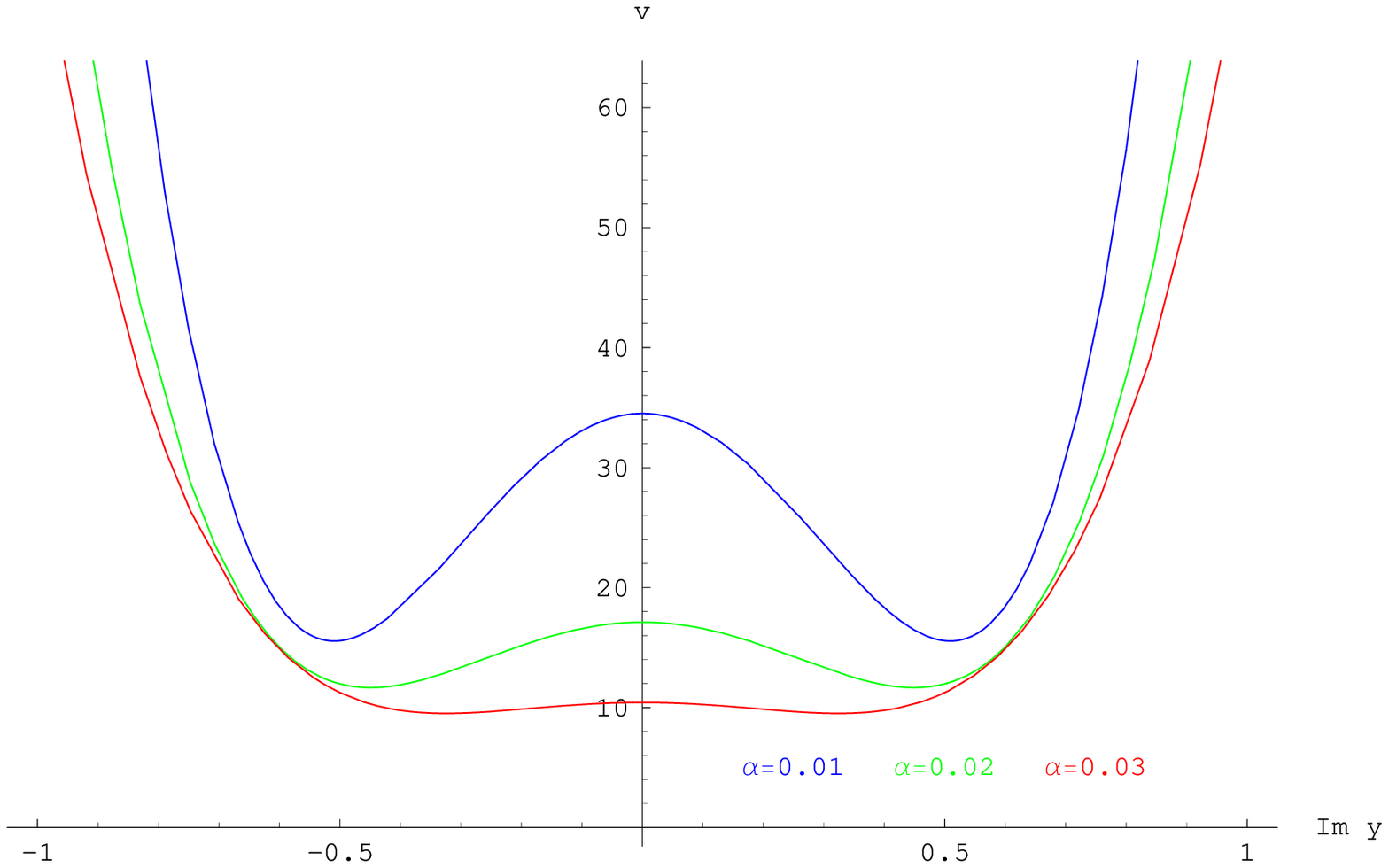} \\
Sections $\Re y=0$ of the scoop branch &  \\
(admitting a classical limit) & Sections $\Im y=const$ of the hump branch \\
& (vanishing in the classical limit)
\end{tabular}
\caption{Different sections of the two branches of $V_{BH}$}
\label{BPSEvol}
\end{figure}

Increasing the value of~$\alpha $ the flatness of the humps increases (on
the figure~\ref{BPSEvol} are presented sections of hump chump by planes~$\Re
y=const$ passing through minima),
and at the critical value~$\alpha _{cr_{1}}$ the humps disappear. For $%
\alpha >\alpha _{cr_{1}}$ the non-BPS branch of $V_{BH}$ in the BPS domain
does not contain critical points of $V_{BH}$ any more.

Let us now analyze the evolution of the BPS branch of $V_{BH}$ when
increasing~$\alpha $. When~$\alpha \rightarrow \frac{1}{3\sqrt{3}}^{-}$, the
BPS minimum point gets closer and closer to the boundary of the region where
$g<0$ (see Fig.~\ref{BPSregions}), until it sits exactly on the curve where
the metric function vanishes, becoming a non-admissible critical point of $%
V_{BH}$ (red curve on the Fig.~\ref{BPSEvol}). Differently from what happens
in the non-BPS branch when $\alpha $ gets bigger than $\alpha _{cr_{1}}$, in
the BPS branch when $\alpha $ gets bigger than the critical value $\frac{1}{3%
\sqrt{3}}$ the critical point does not disappear, but rather it changes its
supersymmetry-preserving features: from BPS it \textit{``transmutes''} into
a non-BPS one, preserving its axion-free character. Such a phenomenon of
\textit{``transmutation'' of attractors} can be seen by looking at the ~$%
q_{0}<0$ branch of Fig.~\ref{BPSSolutions}: when $\alpha $ increases and it
passes through $\frac{1}{3\sqrt{3}}$ the BPS minimum transforms into a
non-BPS minimum.

\begin{figure}[h]
\includegraphics[width=0.9\textwidth]{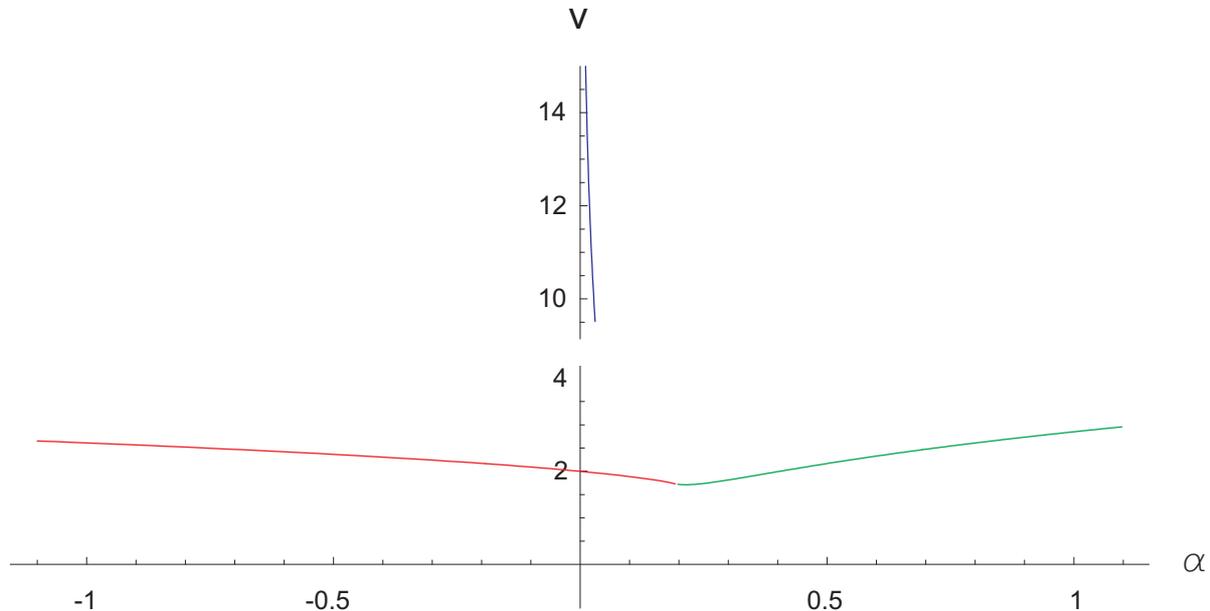}
\caption{$\protect\alpha $-dependence of the BH entropy $S_{BH}$ for~$%
q_{0}<0 $}
\label{EntropyBPS}
\end{figure}

Finally, it is worth plotting in Fig. \ref{EntropyBPS} the $\alpha $%
-dependence of the BPS and non-BPS entropies (e.g. for $q_{0}<0$, without
any loss of generality). Consistently with Fig. \ref{BPSSolutions}, the red
line corresponds to BPS minima, the blue line to non-BPS minima with
non-vanishing axion, and the green line to axion-free non-BPS minima. As it
is seen, as previously mentioned in all the range of $\alpha $ supporting
coexistence of minima it holds $S_{BH,BPS}<S_{BH,non-BPS}$.

Let us also notice that the definitions (\ref{def-BPS}) are dimensionless in
charges, and thus all considered phenomena hold for any magnitude of BH
charges $q_{0}$ and $p^{1}$.

\subsection{\label{Quantum-non-BPS}Non-BPS domain}

\begin{figure}[h]
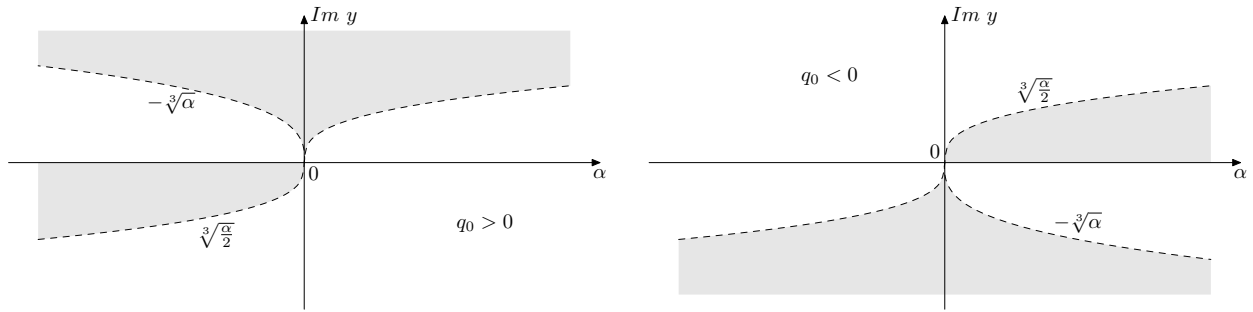

$\epsfxsize=0.45\textwidth\epsfbox{allowedRegionNonBPS.1}$ \quad $\epsfxsize%
=0.45\textwidth\epsfbox{allowedRegionNonBPS.2}$%
\caption{Domains of positivity of $g$ and $e^{K}$}
\label{nonBPSregions}
\end{figure}

Let us switch to a new special coordinate~$y$ and to a new quantum parameter~%
$\alpha $ (notice the different definition with respect to the corresponding
quantities in the BPS domain defined by Eq. (\ref{def-BPS})):
\begin{equation}
t\equiv yp^{1}\sqrt{-\frac{q_{0}}{(p^{1})^{3}}},\qquad \lambda \equiv \alpha
q_{0}\sqrt{-\frac{q_{0}}{(p^{1})^{3}}},  \label{def-non-BPS}
\end{equation}
in terms of which the dependence of all the considered quantities on the
charges is factorized:
\begin{equation}
\begin{array}{l}
\displaystyle W=q_{0}\left( 1+3y^{2}\right) ,\qquad e^{-K}=-4q_{0}\sqrt{-%
\frac{q_{0}}{(p^{1})^{3}}}\left( \alpha -2\Im y^{3}\right) ; \\[0.5em]
\displaystyle g=-\frac{3p^{1}}{q_{0}}\,\frac{\left( \Im y^{3}+\alpha \right)
\Im y}{\left( 2\Im y^{3}-\alpha \right) ^{2}},\qquad V_{BH}=v(y,\bar{y}%
,\alpha )\frac{q_{0}}{\displaystyle\sqrt{-\frac{q_{0}}{(p^{1})^{3}}}}; \\%
[0.5em]
\displaystyle v(y,\bar{y},\alpha )\equiv -\frac{1}{4\Im y\left( \alpha
^{2}-\alpha \Im y^{3}-2\Im y^{6}\right) }\left[ \rule[-1em]{0pt}{2.5em}%
12\alpha ^{2}\left( \Im y^{2}+\Re y^{2}\right) \right. \\[0.5em]
\displaystyle\phantom{V_{eff}=4\Im y\left( \alpha^2 \alpha \Im y^3 \right)}%
\left. +4\Im y^{4}\left( 3\Im y^{4}+12\Im y^{2}\Re y^{2}+\left( 1+3\Re
y^{2}\right) ^{2}\right) \right. \\
\displaystyle\phantom{V_{eff}=4\Im y\left( \alpha^2 \alpha \Im y^3 \right)}%
\left. +\alpha \Im y\left( -3\Im y^{4}+\left( 1+3\Re y^{2}\right) ^{2}+6\Im
y^{2}\left( -3+\Re y^{2}\right) \right) \rule[-1em]{0pt}{2.5em}\right] .
\end{array}
\label{WKgNonBPSD}
\end{equation}
Notice that $v(y,\bar{y},\alpha )$ defined in the non-BPS domain by Eq. (\ref
{WKgNonBPSD}) is different from its counterpart defined in the BPS domain by
Eq. (\ref{WKgBPSD}). As pointed out above, the topology of the allowed
regions of $\Im t$ depends on the sign of $\lambda $, as given by Fig.~\ref
{regions}. In terms of $y$ and $\alpha $ defined in the non-BPS domain by
Eq. (\ref{def-non-BPS}), the corresponding allowed regions are given by Fig.~%
\ref{nonBPSregions}.

\begin{figure}[h]
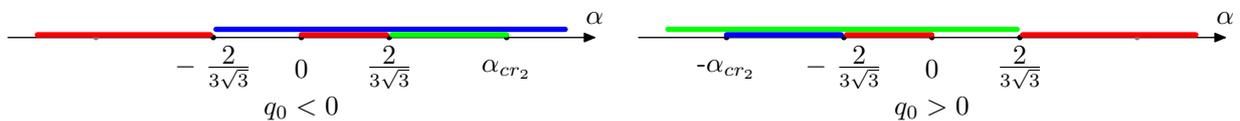

\begin{tabular}{lr}
$\epsfxsize=0.45\textwidth\epsfbox{solutions.3}$ & $\epsfxsize=0.45\textwidth%
\epsfbox{solutions.4}$%
\end{tabular}
\caption{Ranges of the quantum parameter $\protect\alpha $ supporting minima
of $V_{BH}$, for the cases $q_{0}<0$ and $q_{0}>0$. The red line corresponds
to BPS minima, and the green and blue lines to the two different kinds of
axion-free non-BPS minima}
\label{NonBPSSolutions}
\end{figure}

As in the BPS domain, also in the non-BPS domain the phenomena of \textit{%
``separation'' and ``transmutation'' of attractors} arise, but in a
slightly different way. A pictorial view of the ranges of $\alpha $
supporting the various typologies of minima of $V_{BH}$ is given by
Fig.~\ref {NonBPSSolutions}. Differently from what happens in the
BPS domain, in the non-BPS domain no non-BPS minima with
non-vanishing axion exist at all, and there are two distinct
typologies of axion-free non-BPS minima.

Concerning the BPS minima, for $q_{0}<0$ they exist in the range~$0<\alpha <%
\frac{2}{3\sqrt{3}}$, whereas for $q_{0}>0$ they exist for~$\alpha >\frac{2}{%
3\sqrt{3}}$. The same holds for~$\alpha <0$, but with opposite sign of~$%
q_{0} $.

As mentioned, in the considered non-BPS domain there are two kinds of
non-BPS minima, both axion-free (see Fig.~\ref{NonBPSSolutions}). The
corresponding relevant critical value of the quantum parameter is~$\alpha
_{cr_{2}}\approx 0.934$.

As mentioned above, since the BPS critical points can be computed
analytically, one can calculate the $\alpha $-dependent expression of the
BPS BH entropy in the non-BPS domain to be
\begin{equation}
S=\pm \frac{\pi }{4}\frac{\left( 1-3\Im y^{2}\right) ^{2}}{\alpha -2\Im y^{3}%
},\quad \mbox{with $\Im y$ satisfying}\quad \Im y^{3}+\Im y-2\alpha =0,
\end{equation}
where the solution of the cubic equation must be chosen inside the allowed
region(s) of the moduli space, and the $\pm $ branches of $S_{BH}$ must be
chosen in order to obtain $S_{BH}>0$. As it has to be in the non-BPS domain,
in the classical limit the cubic equation has no admissible solutions.

Let us now analyze the evolution $V_{BH}$ with respect to the quantum
parameter $\alpha $, choosing $q_{0}<0$ without loss of generality.

\begin{figure}[h]
\includegraphics[width=0.45\textwidth]{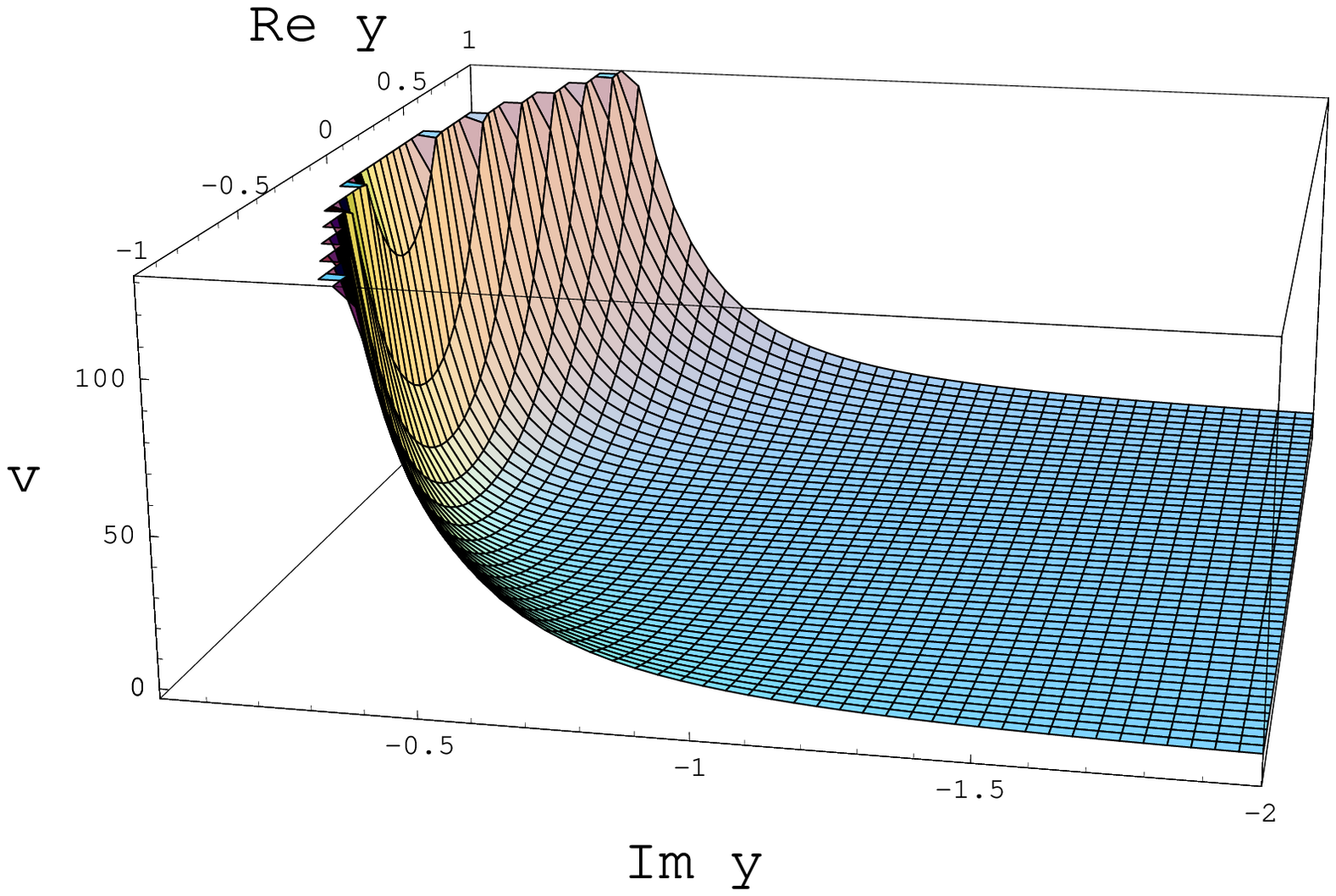} %
\includegraphics[width=0.45\textwidth]{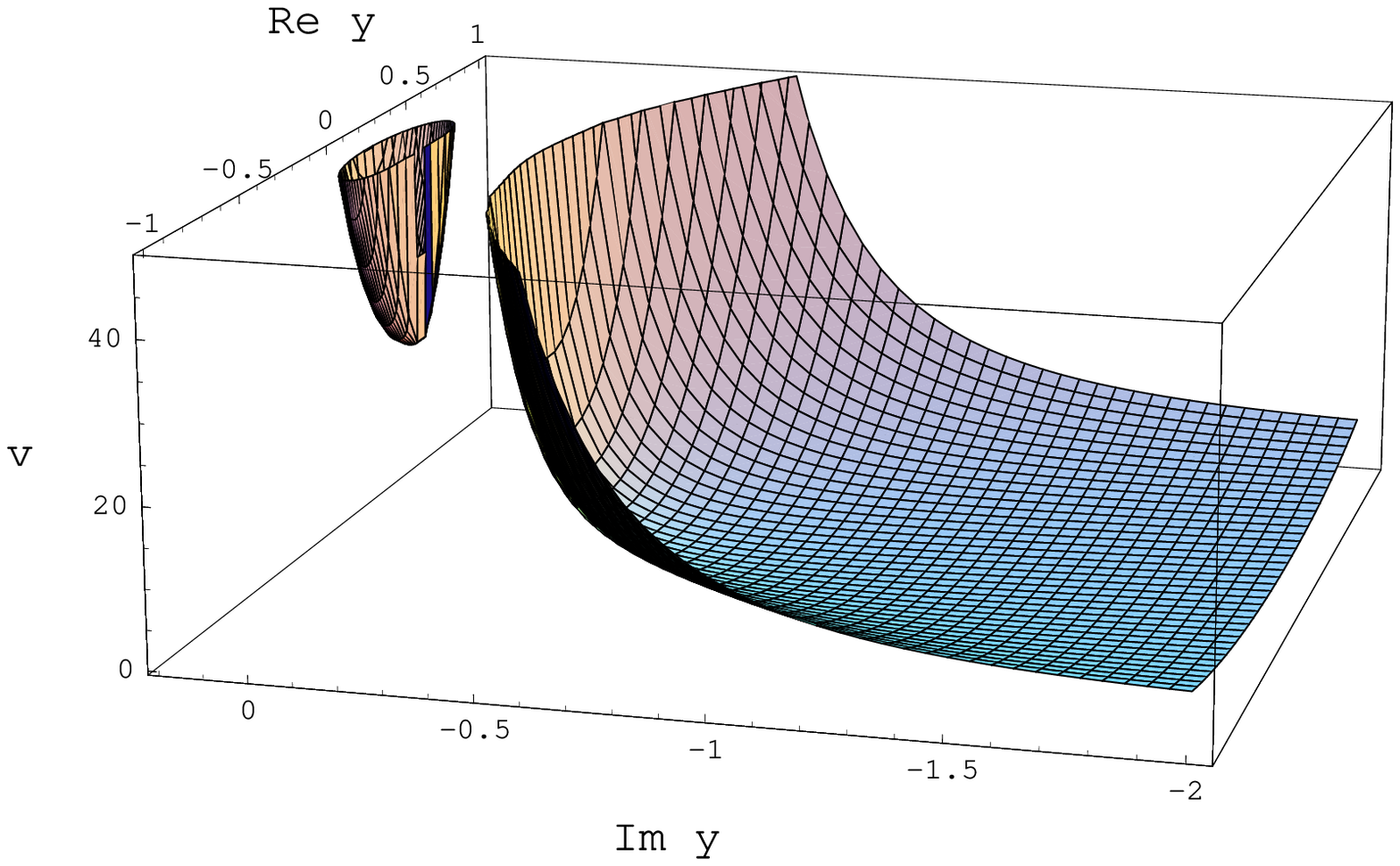}
\caption{Plots of $V_{BH}$ in the classical limit~($\protect\alpha =0$,
left) and~for $\protect\alpha =1/100$ (right)}
\label{NonBPSPlot}
\end{figure}

The classical limit of the function~$v$ defined in Eq. (\ref{WKgNonBPSD})
has the form of a ``scoop'', with a minimum at the point~$\Re y=0$, $\Im y=1$
(see Eq. (\ref{classical-solutions}) and Fig.~\ref{NonBPSPlot}).

As done in the BPS domain, let us now slightly increase the value of $\alpha
$ from $0$ to $0.01$. As one can see from Fig.~\ref{NonBPSPlot}, $\alpha
=0.01$ is in the range supporting the \textit{``separation'' of attractors},%
\textit{\ }in this case corresponding to the coexistence of a BPS attractors
with a non-BPS one, both axion-free. Such stable critical points of $V_{BH}$
respectively have coordinates
\begin{equation}
\begin{tabular}{llll}
$non-BPS:$ & $\Re y=0$, & $\Im y\approx -1.02$, & $V_{BH,non-BPS}\approx 2.03%
\sqrt{-q_{0}(p^{1})^{3}};$ \\[0.3em]
$BPS:$ & $\Re y=0$, & $\Im y\approx 0.02$, & $V_{BH,BPS}\approx 24.98\sqrt{%
-q_{0}(p^{1})^{3}}.$%
\end{tabular}
\label{comparison-2}
\end{equation}

As one can see from the plot of $V_{BH}$ and analogously to what happens in
the BPS domain, in contrast to the classical case there appears a new chump,
which contains a stable BPS critical point (not existing in the classical
limit). The width of such a disconnected branch of $V_{BH}$ (which vanishes
in the classical limit) and its separation from the other branch admitting a
consistent classical limit are both~proportional to $\sqrt[3]{\alpha }$. As
in the BPS domain, such two branches of $V_{BH}$ are separated by a region
in which $g<0$.

As yielded by Eq. (\ref{comparison-2}), for $\alpha =0.01$ it holds that $%
V_{BH,non-BPS}<V_{BH,BPS}$ (see discussion below).

However in the non-BPS domain, once again differently from the BPS domain,
in the range $0<\alpha <\alpha _{cr_{2}}$ supporting the \textit{%
``separation'' of attractors }one can observe also the phenomenon of
the \textit{``transmutation'' }of the supersymmetry-preserving
features of the such minima of $V_{BH}$. This can be seen by
increasing $\alpha $.
\begin{figure}[h]
\includegraphics[width=0.5\textwidth]{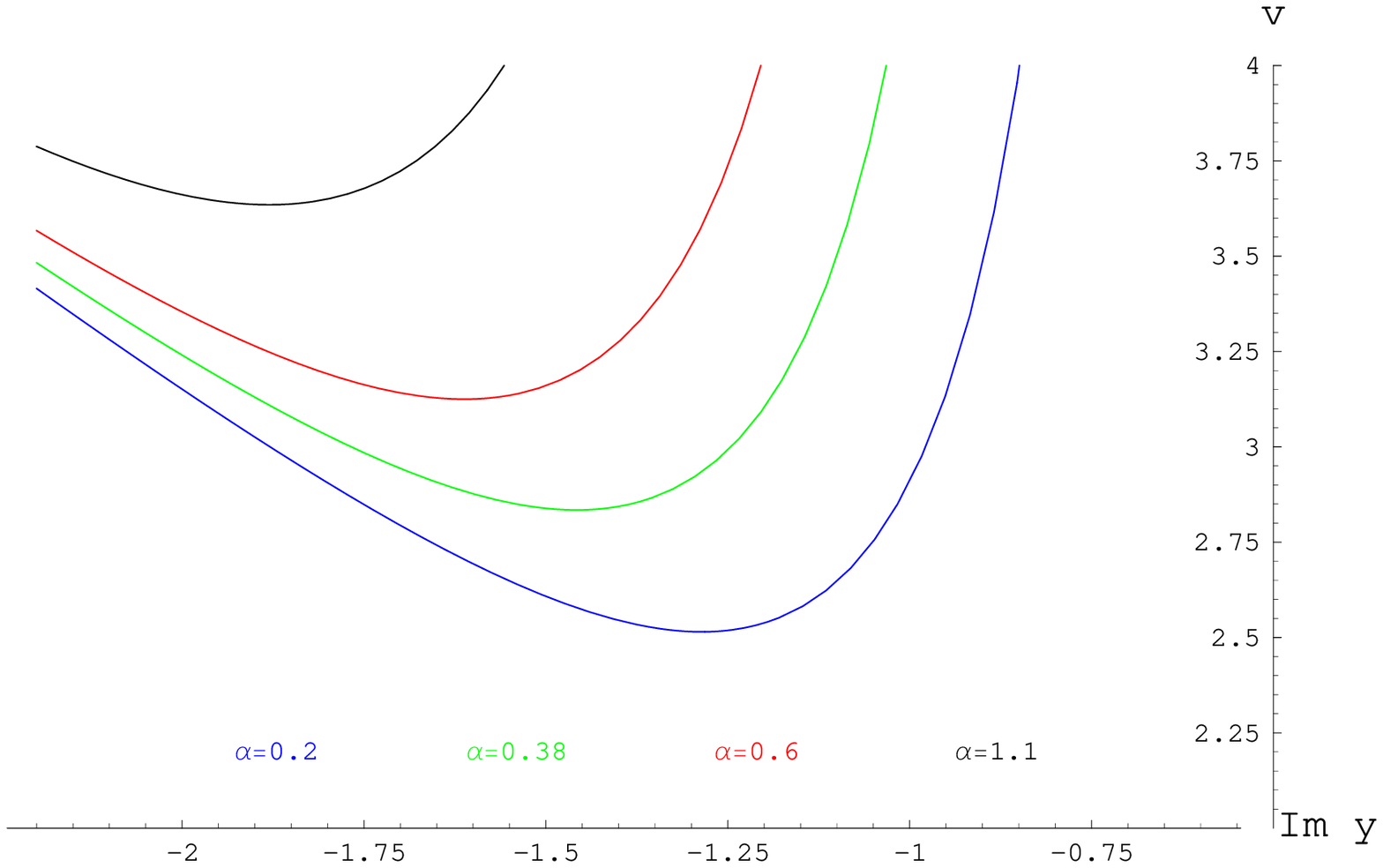} %
\includegraphics[width=0.5\textwidth]{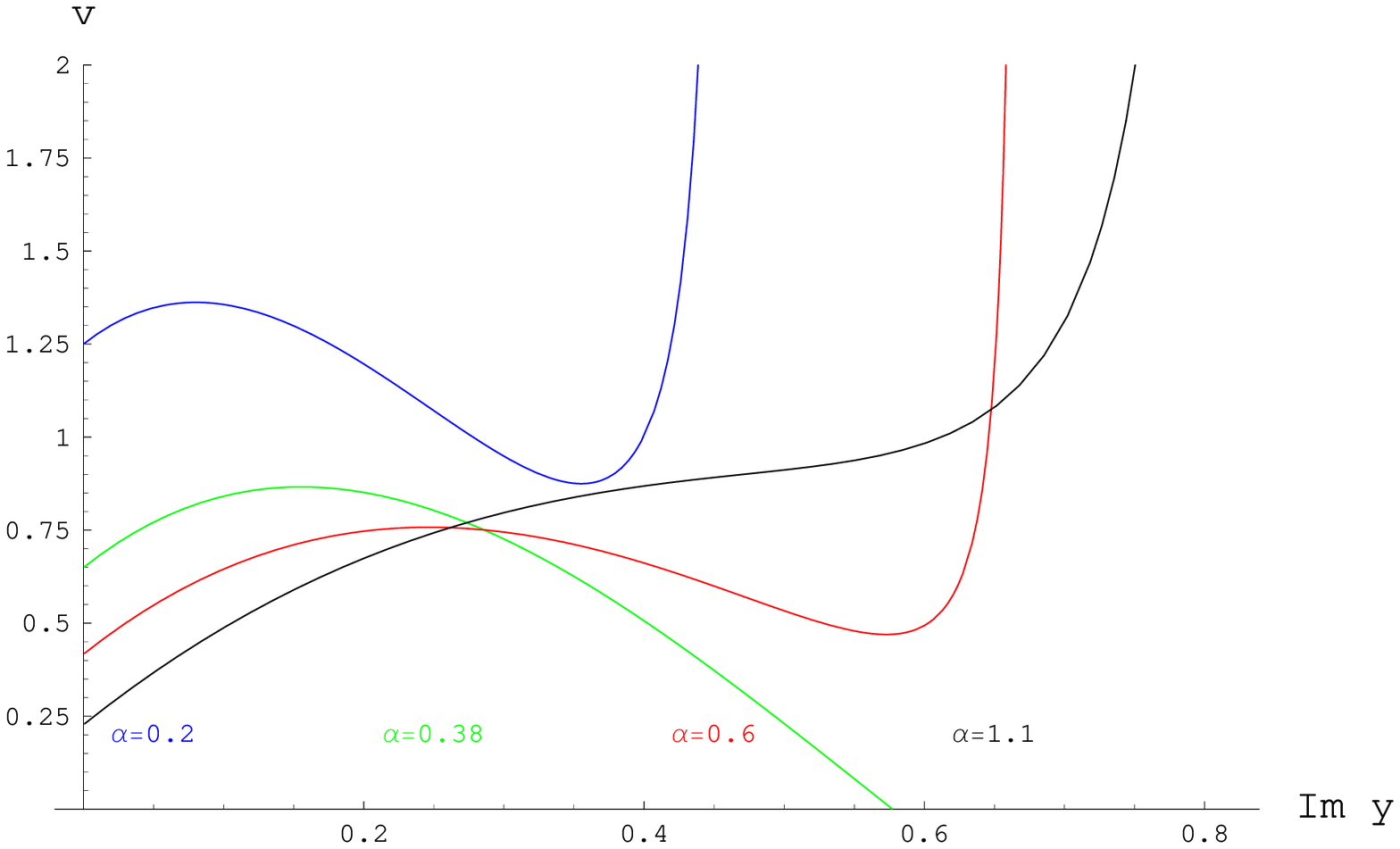}
\caption{The sections~$\Re y=0$ of the two branches of $V_{BH}$: scoop
(admitting a classical limit, left) and hump (vanishing in the classical
limit, right)}
\label{VBHNonBPSDomEvol}
\end{figure}
As evident by looking at Fig.~\ref{VBHNonBPSDomEvol}, the scoop chump of $%
V_{BH}$ (\textit{i.e.} the branch of $V_{BH}$ containing an axion-free
non-BPS minimum and admitting a classical limit), does not undergo any
qualitative change when increasing $\alpha $. On the other hand, the hump
chump of $V_{BH}$ (\textit{i.e.} the one vanishing for $\alpha \rightarrow 0$%
) exhibits a \textit{``transmutation''} of the minimum: for $0<\alpha <\frac{%
2}{3\sqrt{3}}$ it constains a BPS minimum. For~$\alpha =\frac{2}{3\sqrt{3}}$
such a critical point of $V_{BH}$ sits on the boundary (see Fig.~\ref
{nonBPSregions}) where~$e^{K}$ diverges, so it is not an allowed critical
point of $V_{BH}$. For~$\frac{2}{3\sqrt{3}}<\alpha <\alpha _{cr_{2}}$ the
minimum is still present, it is not BPS any more, but rather it is non-BPS
(and axion-free). Thence, for $\alpha >\alpha _{cr_{2}}$ the hump chump of $%
V_{BH}$ does not contain any critical point.

Also the scoop chump of $V_{BH}$ exhibits a \textit{``transmutation''} of
the minimum, but for negative values of $\alpha $. Indeed, when $\alpha $
gets smaller than $-\frac{2}{3\sqrt{3}}$ the axion-free non-BPS minimum
\textit{``transmutes''} into a BPS one (axion-free, as all the admissible
BPS critical points found in our analysis).

The shape of $V_{BH}$ for various values of $\alpha $ is plotted in Fig.~\ref
{VBHNonBPSDomEvol}. One might see that for small values of~$\alpha $ (blue
line) the hump chump contains a BPS minimum. At the critical value $\alpha =%
\frac{2}{3\sqrt{3}}$ (green line) the minimum disappears and
reappears, but non-BPS, when~$\alpha >\frac{2}{3\sqrt{3}}$ (red
line). For $\alpha >\alpha _{cr_{2}}$, no critical points exist at
all (black line).

\begin{figure}[h]
\includegraphics[width=0.6\textwidth]{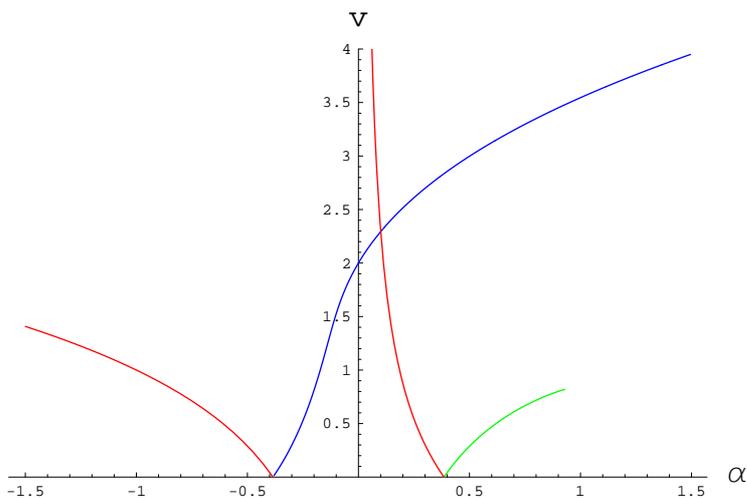}
\caption{$\protect\alpha $-dependence of the BH entropy $S_{BH}$ for~$%
q_{0}<0 $}
\label{EntropyNonBPS}
\end{figure}

The $\alpha $-dependence of the BH entropy is shown in Fig.~\ref
{EntropyNonBPS}. Consistently with the colors used in Fig. \ref
{NonBPSSolutions}, the red plot corresponds to BPS BH entropy, whereas the
blue and green plots to non-BPS BH entropy.

As done in the BPS domain, it is worth pointing out that the definitions (%
\ref{def-non-BPS}) are dimensionless in charges, and thus all considered
phenomena hold for any magnitude of BH charges $q_{0}$ and $p^{1}$.

Thus, the non-BPS domain exhibits various differences with respect to the
BPS domain, which can be summarized as follows:

\textit{i)} all non-BPS minima in such a domain are axion-free;

\textit{ii)} the \textit{``transmutation''} of the supersymmetry-preserving
features of the minimum of $V_{BH}$ happens not only in the branch of $%
V_{BH} $ admitting a classical limit, but also in the one vanishing in such
a limit;

\textit{iii) }the \textit{``transmutation''} inside the branch of $V_{BH}$
vanishing in the classical limit actually determines a \textit{%
``transmutation''} in the region of \textit{``separation''} of
attractors: for $0<\alpha <\frac{2}{3\sqrt{3}}$ a BPS and a non-BPS
minimum coexist; when $\alpha $ gets bigger than
$\frac{2}{3\sqrt{3}}$ the BPS minimum \textit{``transmutes''} into a
non-BPS one, who then coexists with the pre-existing non-BPS
(axion-free) minimum until $\alpha $ reaches the critical value
$\alpha _{cr_{2}}$;

\textit{iv)} as evident by looking at Fig.~\ref{EntropyNonBPS}, in the range
$0<\alpha <\frac{2}{3\sqrt{3}}$ supporting the coexistence of a BPS and a
non-BPS minimum, the sign of $S_{BH,non-BPS}-S_{BH,BPS}$ (for fixed BH
charges $p^{0}$ and $q_{1}$ satisfying $p^{0}q_{1}<0$) changes. Indeed,
there exists a particular value $\widehat{\alpha }\approx 0.1\in \left( 0,%
\frac{2}{3\sqrt{3}}\right) $ for which remarkably $S_{BH,non-BPS}=S_{BH,BPS}$%
. This means that for $\alpha =\widehat{\alpha }$ two \textit{different}
purely charge-dependent attractor configurations of the scalar $t$ exist
(one preserving one half of the $8$ supersymmetries pertaining to the
asymptotical $\mathcal{N}=2$, $d=4$ superPoincar\'{e} algebra and the other
not preserving any of them) such that they determine the \textit{same} BH
entropy. For $0<\alpha <\widehat{\alpha }$ it holds $%
S_{BH,non-BPS}<S_{BH,BPS}$, whereas for $\widehat{\alpha }<\alpha <\frac{2}{3%
\sqrt{3}}$ it holds $S_{BH,BPS}<S_{BH,non-BPS}$;

\textit{v) }for $\frac{2}{3\sqrt{3}}<\alpha <\alpha _{cr_{2}}$ it holds $%
\widetilde{S}_{BH,non-BPS}<S_{BH,non-BPS}$, where $\widetilde{S}%
_{BH,non-BPS} $ is the BH entropy determined by the non-BPS minimum
originated from the BPS one by \textit{``transmutation'' }(see the green and
blue plots in Fig.~\ref{EntropyNonBPS});

\textit{vi}) for $\alpha >\alpha _{cr_{2}}$, the branch of $V_{BH}$
vanishing in the classical limit actually still contains critical points of $%
V_{BH}$, but they are not admissible and/or they are not stable.

It is worth pointing out that all the obtained results are expressed in
terms of coordinates and quantum parameters \textit{dimensionless} in the BH
charges (\textit{i.e.} depending on \textit{ratios} of BH charges, as given
by Eqs. (\ref{def-BPS}) and (\ref{def-non-BPS})). Thus, \textit{apriori}
they hold for every range of values of the BH charges. Actually, we
disregard the actual quantization of the electric and magnetic charges,
considering $\mathbb{R}^{4}$, rather than the 4-dimensional charge lattice $%
\widehat{\Gamma }$, to be the space spanned by the BH charge vector $Q$. In
other words, all our analysis and the obtained results hold in the
semiclassical regime of large, continuous, real BH charges.

\subsection{\label{D0-D6}The $D0-D6$ BH charge configuration}

From the above analysis of the extremal BH attractors supported by the
``magnetic'' charge configuration $Q=\left( q_{0},0,0,p^{1}\right) $ in the $%
t^{3}+i\lambda $ model, one might be lead to argue that starting from a BH
charge configuration supporting only one class of attractors at the
classical level, at the quantum level one obtains that such a charge
configuration generally supports more than one class of attractors (at least
for a certain range of the quantum parameter $\lambda $).

This is actually not true, and a counterexample is provided by the BH charge
configuration $Q=\left( q_{0},0,p^{0},0\right) $, usually named Kaluza-Klein
BH (in $M$-theory language) \cite{FM} or $D0-D6$ system (in Type IIA
Calabi-Yau compactifications in the language of superstring theory) \cite
{TT,TT2}, for which the superpotential acquires the form
\begin{equation}
W=q_{0}-2i\lambda p^{0}+p^{0}t^{3}.
\end{equation}
It is easy to check that at the classical level such a charge configuration
only supports non-BPS attractors \cite{TT,TT2,CFM1}. Differently from what
happens for the ``magnetic'' charge configuration, this holds also at the
quantum level, $\forall \lambda \in \mathbb{R}$; in other words, the $%
t^{3}+i\lambda $ model does not have admissible BPS critical points of $%
V_{BH}$ supported by the $D0-D6$ charge configuration. This can be easily
realized by looking at the real and imaginary parts of the BPS Attractor
Equation for such a charge configuration, respectively of the following form
(having a smooth classical limit):
\begin{equation}
\begin{array}{ll}
\displaystyle\frac{(\Im t^{3}-\lambda )(\Re t^{2}+\Im t^{2})}{\lambda +2\Im
t^{3}}=0; &  \\[1em]
\displaystyle\frac{6\Im t(2\lambda p^{0}\Re t+p^{0}\Im t^{3}\Re t+\Im
t(q_{0}+p^{0}\Re t^{3}))}{\lambda +2\Im t^{3}}=0. &
\end{array}
\end{equation}
By recalling the expression of the metric function $g$ given by Eq. (\ref
{geom-t^3+ic}), it is clear that all BPS critical points of $V_{BH}$
supported by the $D0-D6$ charge configuration are points where $g=0$, and
therefore they are not admissible.

Also in view of the analysis of \cite{Quantum-N=2}, such a result can be
directly extended to the axion-free BPS critical points of $V_{BH}$ in a $%
\mathit{\lambda }$\textit{-corrected} SK cubic geometry, based on the K\"{a}%
hler gauge-invariant holomorphic prepotential (in a suitable basis of
special symplectic coordinates $t^{i}\equiv \varsigma ^{i}-i\omega ^{i}$, $%
i=1,...,n_{V}$)
\begin{equation}
\mathcal{F}=\frac{1}{3!}d_{ijk}t^{i}t^{j}t^{k}+i\lambda ,\label{d-SKG+ic}
\end{equation}
which is nothing but the $n_{V}$ moduli generalization of Eq. (\ref{t^3+ic}%
). The $\mathit{\lambda }$\textit{-corrected} real metric $g_{ij}$ can be
easily computed from Eq. (\ref{d-SKG+ic}) to be
\begin{equation}
g_{ij}=-\frac{3}{2}\left[ \frac{d_{ij}}{d-3\lambda }-\frac{3}{2}\frac{%
d_{i}d_{j}}{\left( d-3\lambda \right) ^{2}}\right] ,\label{CERN1}
\end{equation}
where $d_{ij}\equiv d_{ijk}\omega ^{k}$, $d_{i}\equiv d_{ijk}\omega
^{j}\omega ^{k}$ and $d\equiv d_{ijk}\omega ^{i}\omega ^{j}\omega ^{k}$.
Notice that the limit $\lambda \rightarrow 0$ consistently yields the
expression of the real metric of a SK cubic geometry, given by Eq. (2.4) of
\cite{CFM1}. In \cite{Quantum-N=2} the axion-free BPS critical points of $%
V_{BH}$ in the SK geometry based on $\mathcal{F}$ given by Eq. (\ref
{d-SKG+ic}) have been computed to be
\begin{equation}
t_{BPS,axion-free}^{i}=\omega _{BPS,axion-free}^{i}=i\frac{p^{i}}{\frak{c}},%
\label{CERN2}
\end{equation}
where the charge-dependent quantity $\frak{c}$ satisfies the third order
algebraic equation
\begin{equation}
2\lambda \frak{c}^{3}-q_{0}\frak{c}^{2}+\frac{1}{3!}d_{ijk}p^{i}p^{j}p^{k}=0.
\end{equation}
It is then clear that the $D0-D6$ BH charge configuration does not support
axion-free BPS critical points of $V_{BH}$, because for $p^{i}=0$ $\forall
i=1,...,n_{V}$ Eqs. (\ref{CERN1}) and (\ref{CERN2}) yield $%
g_{ij,BPS,axion-free}=0$.

Notice that the non-admissibility of the axion-free BPS critical points of $%
V_{BH}$ in the $D0-D6$ BH charge configuration would not be evident by only
looking at the BPS BH entropy determined by axion-free solutions, which reads%
\footnote{%
Such an expression corrects a misprint in Eq. (3.35) of \cite{Quantum-N=2}. }
\begin{equation}
S_{BH,BPS,axion-free}=-2\pi \left( \frak{c}+\frac{\left( p^{0}\right) ^{2}}{%
\frak{c}}\right) \left( q_{0}-\frac{3}{2}\lambda \frak{c}\right) ,
\end{equation}
whose limit (independent on $n_{V}$) in the $D0-D6$ BH charge configuration
is finite (and constrains $\lambda \in \mathbb{R}_{0}^{-}$):
\begin{equation}
S_{BH,BPS,axion-free,D0-D6}=-\frac{\pi }{2}\left( \frac{\left( q_{0}\right)
^{2}}{2\lambda }+2\lambda \left( p^{0}\right) ^{2}\right) .
\end{equation}

\section*{\textbf{Acknowledgments}}

The work of S.B., S.F. and A.S. has been supported in part by the European
Community Human Potential Program under contract MRTN-CT-2004-005104 \textit{%
``Constituents, fundamental forces and symmetries of the universe''}.

The work of S.F. has been supported in part by European Community Human
Potential Program under contract MRTN-CT-2004-503369 \textit{``The quest for
unification: Theory Confronts Experiments''}, in association with INFN
Frascati National Laboratories and by D.O.E. grant DE-FG03-91ER40662, Task C.

The work of A.M. has been supported by a Junior Grant of the \textit{%
``Enrico Fermi''} Center, Rome, in association with INFN Frascati National
Laboratories.

\section*{Appendix}

The complex BPS Attractor Eq. for the quantum $1$-modulus model based on the
prepotential $\mathcal{F}=t^{3}+i\lambda $ has the form
\begin{equation}
\frac{t^{4}{p^{1}}-3{p^{1}}{\bar{t}\,}^{2}t^{2}+2{p^{1}}{\bar{t}\,}^{3}t-{%
q_{0}}{\bar{t}\,}^{2}-q_{0}t^{2}+2{q_{0}}{\bar{t}\,}t+8i\lambda {p^{1}}t}{%
\left( t-\bar{t}\right) ^{3}-8i\lambda }=0.
\end{equation}
By taking its real and imaginary parts, one obtains the following two real
Eqs.:
\begin{equation}
\frac{\Re t(\Im t^{3}-\lambda )}{2\Im t^{3}+\lambda }=0;\qquad \frac{\Im
t(\Im t^{3}p^{1}+3\Im t\Re t^{2}p^{1}+2\lambda p^{1}-\Im tq_{0})}{2\Im
t^{3}+\lambda }=0,  \label{BPS-AEs}
\end{equation}
whose solution reads
\begin{equation}
\Re t=0,\qquad \Im t^{3}p^{1}-\Im tq_{0}+2\lambda p^{1}=0.
\end{equation}
Obviously, the number of real roots of the obtained algebraic cubic Eq.
depends on the relations between its coefficients.

On the other hand, the real part of the complex non-BPS Attractor Eq. for
the model based on $\mathcal{F}=t^{3}+i\lambda $ reads
\begin{equation}
\frac{\Re t[8\Im t^{6}p^{1}-\Im t^{3}\lambda p^{1}+2\lambda ^{2}p^{1}+4\Im
t^{4}(3\Re t^{2}p^{1}-q_{0})+\Im t\lambda (-3\Re t^{2}p^{1}+q_{0})]}{4\Im
t^{7}-2\Im t^{4}\lambda -2\Im t\lambda ^{2}}=0;  \label{Re-non-BPS-AEs}
\end{equation}
its imaginary part is rather cumbersome, and it is not worth presenting it
here.

The above Eqs. get modified when passing to the special coordinate and
quantum parameter suitably defined in the BPS and non-BPS domains (see
definitions (\ref{def-BPS}) and (\ref{def-non-BPS}), respectively). In the
following treatment we will analyze them case-by-case.

\subsection*{BPS domain}

\subsubsection*{BPS Attractor Equations}

In the BPS domain the BPS Attractor Equations (AEs) (\ref{BPS-AEs}) acquire
the form
\begin{equation}
\frac{\left( (\Im y)^{3}-\alpha \right) \Re y}{\alpha +2(\Im y)^{3}}%
=0,\qquad \frac{\Im y\left( 2\alpha +\Im y\left( -1+(\Im y)^{2}+3(\Re
y)^{2}\right) \right) }{\alpha +2(\Im y)^{3}}=0.  \label{BPSEqBPSDom}
\end{equation}
From~Eq. (\ref{WKgBPSD}) it follows that, in order to have a regular
solution, the real part of the modulus~$y$ has to vanish, while its
imaginary part has to satisfy a cubic Eq.:
\begin{equation}
\Re y=0,\qquad (\Im y)^{3}-\Im y+2\alpha =0.  \label{BPSsolBPSD}
\end{equation}
Depending on $\alpha $, the obtained cubic Eq. has up to three real roots.
If~$\alpha ^{2}<1/27$ there are three real roots but, due to the consistency
conditions on the metric and K\"{a}hler potential, only one of them turns
out to determine an admissible critical point of $V_{BH}$. On the other
hand, for~$\alpha ^{2}>1/27$ the situation is described by Fig.~\ref
{BPSSolutions}. Notice that, as it has to be in the BPS domain, in the
classical limit $\alpha \rightarrow 0$ one reobtains the classical solutions
given by Eq. (\ref{classical-solutions}).

\subsubsection*{Non-BPS Attractor Equations}

In the BPS domain the real part (\ref{Re-non-BPS-AEs}) of the non-BPS AEs
acquire the form
\begin{equation}
\frac{\Re y}{\Im y}\,\frac{2\alpha ^{2}-\alpha \Im y(-1+\Im y^{2}+3\Re
y^{2})+4\Im y^{4}(-1+2\Im y^{2}+3\Re y^{2})}{(\alpha -\Im y^{3})(\alpha
+2\Im y^{3})}=0.  \label{ReCritBPSD}
\end{equation}
Once again, the imaginary part is cumbersome, and it is not worth writing it
here. Eq.~(\ref{ReCritBPSD}) can be solved with respect to $\Re y$,
obtaining two solutions:
\begin{equation}
\Re y=0;\qquad \mbox{or}\qquad \Re y^{2}=-\frac{1}{3}\frac{2\alpha
^{2}+\alpha \Im y-\alpha \Im y^{3}-4\Im y^{4}+8\Im y^{6}}{4\Im y^{4}-\alpha
\Im y}.  \label{ReyBPSD}
\end{equation}
We are now going to substitute each of these solutions in the imaginary part
of the non-BPS AEs.

\paragraph*{\textbf{Axion-Free Non-BPS Solutions}}

In this case the imaginary part of the non-BPS AEs acquires form
\begin{equation}
\left( \Im y^{3}-\Im y+2\alpha \right) \left[ 2\alpha ^{3}-5\alpha ^{2}\Im
y-3\alpha ^{2}\Im y^{3}+4\alpha \Im y^{4}+36\alpha \Im y^{6}-8\Im y^{7}-8\Im
y^{9}\right] =0.  \label{12-Oct-1}
\end{equation}
It is worth remarking that the cubic term $\Im y^{3}-\Im y+2\alpha $ which
gets factorized determines the BPS solution~(\ref{BPSsolBPSD}), and thus
here it has to be disregarded. The other factor in Eq. (\ref{12-Oct-1}) is
the one yielding the non-BPS solutions. As it has to be since we are
considering the BPS domain, in the classical limit~$\alpha \rightarrow 0$
this factor reduces to a polynomial of the form~$x^{9}+x^{7}$, which has no
admissible solutions.

Since the solution of an algebraic Eq. of ninth order cannot be found
analytically, one can analyze it numerically, and determine the number of
real solutions depending on the value of $\alpha $. It turns out that~$%
\alpha $ has two critical values: one of them~is $\frac{1}{3\sqrt{3}}$,
while the other only numerically,~$\widetilde{\alpha }_{cr_{1}}\in
(1.1010864,1.1010865)$. For~$0<\alpha <\frac{1}{3\sqrt{3}}$ and $\alpha >%
\widetilde{\alpha }_{cr_{1}}$, only one real root exists, while for~$\frac{1%
}{3\sqrt{3}}<\alpha <\widetilde{\alpha }_{cr_{1}}$ there are tree real
roots. By choosing $q_{0}<0$ (without any loss of generality) and requiring
the stability of the solutions and the belonging to the allowed regions of
the moduli space given by Fig.~\ref{BPSregions}, one gets that in the BPS
domain stable and admissible non-BPS axion-free critical points of $V_{BH}$
exist for~$\alpha >\frac{1}{3\sqrt{3}}$ (see the green line in the plot on
the left of Fig. \ref{BPSSolutions}).

\paragraph*{Non-\textbf{Axion-Free Non-BPS Solutions}}

The other possibility is to substitute the more complicated expression for $%
\Re y$ given by Eq.~(\ref{ReyBPSD}) into the imaginary part of the non-BPS
AEs. By doing so, the imaginary part of the non-BPS AEs reduces to an
algebraic Eq. of fifteenth order:
\begin{equation}
\begin{array}{l}
\displaystyle2\alpha ^{5}+\alpha ^{4}\Im y-13\alpha ^{4}\Im y^{3}+8\alpha
^{3}\Im y^{4}-76\alpha ^{3}\Im y^{6}-120\alpha ^{2}\Im y^{7}- \\[0.5em]
\displaystyle\phantom{2\alpha^5 +}-92\alpha ^{2}\Im y^{9}+320\alpha \Im
y^{10}-32\alpha \Im y^{12}-128\Im y^{13}-32\Im y^{15}=0,
\end{array}
\label{Oct-12-evening-2}
\end{equation}
which clearly cannot be solved analytically. As it has to be since we are
considering the BPS domain, in the classical limit~$\alpha \rightarrow 0$
Eq. (\ref{Oct-12-evening-2}) reduces to a polynomial of the form~$%
x^{15}+x^{13}$, which has no admissible solutions.

By inspecting the number of real roots of such an Eq. with respect to the
value of $\alpha $, one obtains two critical values of $\alpha $. One of
them is known just numerically:~$\alpha _{cr_{1}}\in \left(
0.030101,0.030102\right) $, while the other~one is~$\frac{2}{3\sqrt{3}}$.
When~$0<\alpha <\alpha _{cr_{1}}$, five real roots exist but only one of
them is admissible, whereas~for $\alpha _{cr_{1}}<\alpha <\frac{2}{3\sqrt{3}}
$ and~$\alpha >\frac{2}{3\sqrt{3}}$ three and one real root exist
respectively, but none of them is admissible. Thus (by choosing $q_{0}<0$,
once again without loss of generality), in the BPS domain stable and
admissible non-BPS critical points of $V_{BH}$ with non-vanishing axion
exist for $0<\alpha <\alpha _{cr_{1}}$ (see the blue line in the plot on the
left of Fig. \ref{BPSSolutions}).

\subsection*{Non-BPS domain}

\subsubsection*{BPS Attractor Equations}

In the non-BPS domain the BPS AEs (\ref{BPS-AEs}) acquire the form
\begin{equation}
\frac{\left( \Im y^{3}+\alpha \right) \Re y}{\alpha -2\Im y^{3}}=0,\qquad
\frac{\Im y\left( -2\alpha +\Im y\left( 1+\Im y^{2}+3\Re y^{2}\right)
\right) }{\alpha -2\Im y^{3}}=0.  \label{BPSEq-NBPSDom}
\end{equation}
From~Eq. (\ref{WKgNonBPSD}) it follows that, in order to have a regular
solution, the real part of the modulus~$y$ has to vanish, while its
imaginary part has to satisfy a cubic Eq.:
\begin{equation}
\Re y=0,\qquad \Im y^{3}+\Im y-2\alpha =0.
\end{equation}
Such a cubic Eq. has only one real root $\forall \alpha \in \mathbb{R}$:
\begin{equation}
\Im y=\sqrt[3]{\alpha -\sqrt{\alpha ^{2}+\frac{1}{27}}}+\sqrt[3]{\alpha +%
\sqrt{\alpha ^{2}+\frac{1}{27}}}.  \label{12-Oct-2}
\end{equation}
Even if such a solution exists $\forall \alpha $, the admissible BPS minima
exist for $q_{0}<0$ in the range~$0<\alpha <\frac{2}{3\sqrt{3}}$, and for $%
q_{0}>0$ in the range $\alpha >\frac{2}{3\sqrt{3}}$. The same holds for~$%
\alpha <0$, but with opposite sign of~$q_{0}$ (see the red lines in Fig. \ref
{NonBPSSolutions}). Notice that, as it has to be in the non-BPS domain, in
the classical limit $\alpha \rightarrow 0$ the cubic Eq. has not admissible
solutions.

\subsubsection*{Non-BPS Attractor Equations}

In the non-BPS domain the real part (\ref{Re-non-BPS-AEs}) of the non-BPS
AEs acquire the form
\begin{equation}
\frac{\Re y}{\Im y}\,\frac{2\alpha ^{2}+\alpha \Im y(1+\Im y^{2}+3\Re
y^{2})+4\Im y^{4}(1+2\Im y^{2}+3\Re y^{2})}{(\alpha +\Im y^{3})(\alpha -2\Im
y^{3})}=0.  \label{ReCritNBPSD}
\end{equation}
Once again, the imaginary part is cumbersome, and it is not worth writing it
here. Eq.~(\ref{ReCritNBPSD}) can be solved with respect to $\Re y$,
obtaining two solutions:
\begin{equation}
\Re y=0;\qquad \mbox{either}\qquad \Re y^{2}=-\frac{1}{3}\frac{2\alpha
^{2}+\alpha \Im y+\alpha \Im y^{3}+4\Im y^{4}+8\Im y^{6}}{4\Im y^{4}+\alpha
\Im y}.  \label{12-Oct-evening-1}
\end{equation}
Notice that the non-axion-free branch of solutions given by the second
expression of Eq. (\ref{12-Oct-evening-1}) is not consistent with the
classical admissible and stable non-BPS solutions, which are axion-free (see
Eq. (\ref{classical-solutions}); on the other hand, as stated below,
admissible and stable non-axion-free non-BPS solutions do not exist in the
considered quantum case, too).

We are now going to substitute each of these solutions in the imaginary part
of the non-BPS AEs.

\paragraph*{Axion-Free Non-BPS Solutions}

In this case the imaginary part of the non-BPS AEs acquires form
\begin{equation}
\left( \Im y^{3}+\Im y-2\alpha \right) \left[ 2\alpha ^{3}-5\alpha ^{2}\Im
y+3\alpha ^{2}\Im y^{3}-4\alpha \Im y^{4}+36\alpha \Im y^{6}-8\Im y^{7}+8\Im
y^{9}\right] =0.  \label{12-Oct-3}
\end{equation}
It is worth remarking that the cubic term $\Im y^{3}+\Im y-2\alpha $ which
gets factorized determines the BPS solution~(\ref{12-Oct-2}), and thus here
it has to be disregarded. The other factor in Eq. (\ref{12-Oct-3}) is the
one yielding the axion-free non-BPS solutions, and, as it has to be in the
non-BPS domain, in the classical limit $\alpha \rightarrow 0$ it gives the
classical solutions (see Eq. (\ref{classical-solutions})).

Since the solution of an algebraic Eq. of ninth order cannot be found
analytically, as done in the BPS domain one can analyze it numerically, and
determine the number of real solutions depending on the value of $\alpha $.
It turns out that~$\alpha $ has two relevant critical values, one is $\frac{2%
}{3\sqrt{3}}$ and the other is known only numerically:~$\alpha
_{cr_{2}}\approx 0.934$. For~$\alpha >\alpha _{cr_{2}}$ and $\alpha <0$,
only one real root exists, while for~$0<\alpha <\alpha _{cr_{2}}$ there are
tree real roots. By choosing $q_{0}<0$ (without any loss of generality) and
requiring the stability of the solutions and the belonging to the allowed
regions of the moduli space given by Fig.~\ref{nonBPSregions}, one gets that
in the non-BPS domain stable and admissible non-BPS axion-free critical
points of $V_{BH}$ exist for~$\alpha >-\frac{2}{3\sqrt{3}}$ and that they
double in the range $\frac{2}{3\sqrt{3}}<\alpha <\alpha _{cr_{2}}$ (see the
blue and green lines in the plot on the left of Fig. \ref{NonBPSSolutions}).

\paragraph*{Non-Axion-Free Non-BPS Solutions}

The other possibility is to substitute the more complicated expression for $%
\Re y$ given by Eq.~(\ref{12-Oct-evening-1}) into the imaginary part of the
non-BPS AEs. By doing so, as in the BPS domain the imaginary part of the
non-BPS AEs reduces to an algebraic Eq. of fifteenth order:
\begin{equation}
\begin{array}{l}
\displaystyle2\alpha ^{5}+\alpha ^{4}\Im y+13\alpha ^{4}\Im y^{3}-8\alpha
^{3}\Im y^{4}-76\alpha ^{3}\Im y^{6}-120\alpha ^{2}\Im y^{7}- \\[0.5em]
\displaystyle\phantom{2\alpha^5 +}+92\alpha ^{2}\Im y^{9}-320\alpha \Im
y^{10}-32\alpha \Im y^{12}-128\Im y^{13}+32\Im y^{15}=0,
\end{array}
\end{equation}
which cannot be solved analytically. By performing a numerical inspection of
the number of real roots depending on the values of $\alpha $, and imposing
the conditions of stability and belonging to the allowed regions of the
moduli space given by Fig.~\ref{nonBPSregions}, in this case one finds that
there are no admissible solutions. In other words, in the non-BPS domain
admissible and stable non-BPS critical points of $V_{BH}$ with non-vanishing
axion do not exist $\forall \alpha \in \mathbb{R}$.

\end{document}